\documentclass[english,runningheads]{tlp}
\usepackage{subcaption} 

\usepackage[explicit]{titlesec}  

\usepackage[latin1]{inputenc}
\usepackage[T1]{fontenc}
\usepackage{babel}
\usepackage{graphicx}
\usepackage{amsmath}
\usepackage{amsfonts}
\usepackage{amssymb}    %
\usepackage{multirow}
\usepackage{multicol}
\usepackage{breakcites}
\usepackage{url}
\usepackage{stmaryrd}
\usepackage[nomath]{lmodern}  
\usepackage{longtable}
\usepackage{rotating}
\usepackage{array}
\usepackage{pdflscape}
\usepackage{soul}
\usepackage{xspace}
\usepackage{enumitem}
\usepackage{hyperref}
\usepackage{color}
\usepackage{wrapfig}  
\usepackage{colortbl} 
\usepackage{fix-cm}  

\usepackage{proof}
\usepackage{makecell}
\usepackage{listings}

\renewcommand{\cite}{\citep}

\newcommand{\tts}[1]{{\small{\texttt{#1}}}}
\newcommand{\ttsi}[1]{{\small{\textsf{\textit{{#1}}}}}}

\newcommand{\Cata}{${\mathcal T}_{\mathit{cata}}$}

\newenvironment{sizepar}[2]
{\par\fontsize{#1}{#2}\selectfont}
{\par}

\newcommand{\If}{\leftarrow}

\newcommand{\plus}{\!+\!}

\newtheorem{example}{Example}   
\newtheorem{definition}{Definition}   
\newtheorem{theorem}{Theorem}   

\begin{document}

	\lefttitle{Emanuele De Angelis, Fabio Fioravanti, Alberto Pettorossi, Maurizio Proietti}

	\jnlPage{1}{x}
	\jnlDoiYr{2021}
	\doival{10.1017/xxxxx}

	\title{Verifying Catamorphism-Based Contracts using Constrained Horn Clauses
	}

	\begin{authgrp}

		\author{\hspace*{-3mm}\centering{\sn{Emanuele} \gn{De Angelis} \hspace*{25.5mm} \sn{Fabio} \gn{Fioravanti}}}
		\affiliation{\hspace*{19mm}\centering{IASI-CNR, Rome, Italy \hspace*{17mm} DEc, University of Chieti-Pescara, Pescara, Italy}}

		\author{ \centering{\sn{Alberto} \gn{Pettorossi}  \hspace*{26mm} \sn{Maurizio} \gn{Proietti}}}
		\affiliation{\centering{\hspace*{-15mm}DICII, University of Rome `Tor Vergata', Rome, Italy \hspace*{12mm}IASI-CNR, Rome, Italy}}

	\end{authgrp}

	\maketitle
	\vspace*{-5mm}            





	\begin{abstract}
		We address the problem of verifying that the functions of a program meet
their contracts, specified by pre/postconditions.
We follow an approach based on
{\em constrained Horn clauses} (CHCs) by which
the verification problem is reduced to the problem of checking  satisfiability
of a set of clauses derived from the given program and
contracts.
We consider programs that manipulate {\em algebraic data types} (ADTs)
and a class of contracts specified by 
{\rm catamorphisms}, that is, functions defined by 
simple recursion schemata
on the given ADTs.
We show by several examples that state-of-the-art CHC satisfiability tools
are not effective at solving
the satisfiability problems obtained {by direct translation of
the contracts into CHCs.}
To overcome this difficulty, 
we propose a transformation technique that removes the ADT terms
from CHCs and derives new sets of clauses
that work  on basic {sorts} only, such as integers and booleans.
Thus, when using the derived CHCs there is no need for induction rules on ADTs.
We prove that the transformation is sound, that is, 
if the derived set of CHCs is satisfiable,
then so is the original set.
We also prove that the transformation always terminates for the class
of contracts specified by catamorphisms.
Finally, we present the experimental results obtained by an implementation of
our technique when verifying
many non-trivial contracts for ADT manipulating programs.

\smallskip
\noindent
\textit{Under consideration for acceptance in TPLP}.
\vspace{-2mm}

	\end{abstract}
	\noindent
	\hrulefill
	\vspace*{-2mm}


	\section{Introduction} 

	\label{sec:Intro}
	\vspace*{-1mm}

Many program verification techniques are based on the classical
{\rm axiomatic} approach proposed by 
Hoare~\citeyearpar{Hoa69}, 
where the functional correctness of a program is specified by
a pair of assertions of first order logic: 
a {\em precondition}, which is assumed to
hold on the program variables before execution, and a {\em postcondition},
which is expected to hold after execution.
This pair of assertions
is often referred to as a {\em contract}~\cite{Meyer92}, and
many programming languages provide built-in support for contracts
associated with function definitions 
(see, for instance,
Ada~\cite{BoochB94}, Ciao~\cite{Hermenegildo&12}, 
and Scala~\cite{OderskySV11}).
In order to prove that all program functions meet their contracts,
{\em program verifiers} generate verification conditions, 
that is, formulas of first order logic 
that have to be
discharged by a theorem prover.
Recent developments of Satisfiability Modulo Theory (SMT)
solvers~\cite{DeB08,CVC4,Ko&14,HoR18}
provide support for
proving verification conditions 
in a wide range of logical theories that axiomatize data types, such as
booleans, uninterpreted functions, linear integer or real arithmetic,
bit vectors, arrays, strings, algebraic data types, and heaps.
Among the program verifiers that use SMT solvers as a back-end, we mention
{\rm Boogie}~\cite{Boogie}, {\rm Dafny}~\cite{Lei13}, {\rm Leon}~\cite{Su&11},
{\rm Stainless}~\cite{HamzaVK19},  and  {\rm Why3}~\cite{Why3}.
There are, however, various 
issues that remain to be solved 
when following this approach to contract verification.
For programs manipulating ADTs, like lists or trees,
one such issue 
is that the verifier often has to generate suitable loop invariants 
whose verification may require the extension of 
SMT solvers with inductive proof rules~\cite{ReK15}.

An alternative approach is based on
translating the contract verification problem
into an equivalent satisfiability problem for
{\em constrained Horn clauses}\footnote{In recent verification literature, 
the term {\em constrained Horn clauses} is often used 
instead of {\em constraint logic programs}, as the
focus is on their logical meaning rather than their
execution as programs.} (CHCs),
that is, Horn clauses extended with logical theories that axiomatize data types like the ones
mentioned above~\cite{JaM94,Gr&12,Bj&15,DeAngelisFGHPP21}.
For clauses extended with theories on basic sorts, such as the theories of
boolean values and linear integer arithmetic, 
various state-of-the-art CHC solvers are available. Among them, let us
mention Eldarica~\cite{HoR18} and SPACER~\cite{Ko&14} that
are quite effective in checking clause satisfiability.
For clauses defined on ADTs, some solvers that
can handle them, have been recently proposed. They 
are based on the extension of 
the satisfiability algorithms 
by induction 
rules~\cite{Un&17,Ya&19}, tree
automata~\cite{KostyukovMF21}, and abstractions~\cite{GovindSG22}.

In this paper we present a method for proving the satisfiability of 
CHCs defined on ADTs that avoids the need of extending 
the satisfiability algorithms and, instead,
follows a transformational approach~\cite{De&18a,De&22a}.
A set of CHCs is transformed, by applying the fold/unfold rules~\cite{EtG96,TaS84},
into a new set of CHCs such that: (i) the ADT terms are no longer present, and hence no induction rules are needed to reason on them,
and (ii)~the satisfiability of the derived set implies the satisfiability
of the original set.
The transformational approach 
has the advantage of separating the concern of dealing
with ADTs (which we face 
 at transformation time) from the concern of dealing with
simpler, non-inductive constraint theories 
(which we face 
at solving time by applying CHC solvers that support basic sorts only).
We  show that the transformational
approach is well suited for a 
significant class of verification problems
where program contracts are
specified by means of {\em catamorphisms}, that is,
functions defined by a simple
structural recursion schema over 
the ADTs manipulated by the program~\cite{MeijerFP91,SuterDK10}.

The main contributions of this paper are the following.
%
(i)~We define a class of CHCs that represent 
{ADT manipulating programs and their contracts.}
No restrictions are imposed on programs, while contracts 
can be specified by means of catamorphisms only (see Section~\ref{sec:Cata}).
%
(ii)~We define an algorithm that, by making use of the given contract 
specifications as lemmas,  transforms a set of CHCs
into a new set of CHCs {without ADT terms} such that, if the transformed clauses are satisfiable,
so are the original ones, and hence
the contracts specified by the original clauses are valid (see Section~\ref{sec:Strategy}).
%
(iii)~Unlike previous work~\cite{De&18a,De&22a}, we prove that 
the transformation algorithm
{\em terminates} for all sets of CHCs in the given class,
and it introduces 
in a fully automatic way  
new predicates corresponding to loop invariants
(see Section~\ref{sec:Strategy}).
%
(iv)~Finally, by using a prototype implementation of our method, 
we prove many non-trivial contracts relative to programs 
{that manipulate lists and trees}
(see Section~\ref{sec:Experiments}). 
\vspace*{-1mm}


%
%

	\section{Preliminaries on  Constrained Horn Clauses} 
	\label{sec:CHCs}
	We consider CHCs defined in a many-sorted first order language with equality
that includes the language of linear integer arithmetic (\textit{LIA})
and boolean expressions (\textit{Bool\/}). For notions 
not recalled here we refer to the literature~\cite{JaM94,Bj&15}.
A {\em constraint} is a quantifier-free formula $c$, where 
the linear integer constraints may occur as subexpressions of
boolean constraints, according to the SMT approach~\cite{Ba&09}. 
The formula $c$ is constructed as follows:

\smallskip
{
$c
::=
d
\,|\,
\textit{B}
\,|\,
\textit{true}
\,|\,
\textit{false}
\,|\sim\!c
\,|\,
c_1\,\&\,c_2
\,|\,
c_1\!\!\vee\!c_2
\,|\,
c_1\!\Rightarrow\!c_2
\,|\,
c_1\!=\!c_2
\,|\,
\textit{ite}(c,c_1,c_2)
\,|\,
t\!=\!\textit{ite}(c,t_1,t_2)$

$d ::=\, t_1\!\!=\!t_2 \,|\, t_1\!\!\geq\!t_2 \,|\, t_1\!\!>\!t_2 \,|\, t_1\!\!\leq\!t_2 \,|\, t_1\!\!<\!t_2 $
}

\smallskip
\noindent
where 
{$B$} is a boolean variable and
{$t$}, possibly with subscripts,  
is a \textit{LIA} term of the form {$a_0+a_1X_1+\dots+a_nX_n$}
with integer coefficients {$a_0,\dots, a_n$}
and variables
{$X_1,...,X_n$}.
The `$\sim$' symbol denotes negation.
The ternary function \textit{ite} denotes the if-then-else operator.
The equality `=' symbol is used for both integers and booleans.

An {\it atom} is a formula of the form $p(t_{1},\ldots,t_{m})$, where~$p$
is a predicate symbol not occurring in $\textit{LIA}\cup\textit{Bool\/}$,
and $t_{1},\ldots,t_{m}$ are first order terms.
A~{\it constrained Horn clause}  (or a CHC, or simply, a {\it clause}) is
an implication of the form $H\leftarrow c, G$.
The conclusion (or {\it head\/})~$H$ is either an atom or \textit{false},
the premise (or {\it body\/}) is the conjunction of a constraint~$c$
and a (possibly empty) conjunction~$G$ of atoms.
A clause is called a {\it  goal\/} if its head is \textit{false}, and
a {\it definite clause\/}, otherwise.
Without loss of generality, we assume that every atom occurring in the body
of a clause has distinct variables (of any sort) as arguments.
%
By ${\it vars}(e)$ we denote the set of all variables occurring in an
{expression} $e$.
Given a formula $\varphi$, we denote by $\forall (\varphi)$ its
{\it universal closure}.
Let~$\mathbb D$ be the usual interpretation for the symbols of {theory}
$\textit{LIA}\cup\textit{Bool\/}$.
By $M(P)$ we denote the {\it least} ${\mathbb D}$-model of a set $P$ of definite  clauses~\cite{JaM94}.
In the examples, we will use the Prolog syntax and the teletype font.
Moreover, we will often prefer writing \tts{B1} and 
\tts{\textasciitilde B2}, 
instead of the equivalent constraints \tts{B1=true} and \tts{B2=false}, respectively.

\vspace{-4mm}

	\section{A Motivating Example} 
	\label{sec:IntroExample}
	The CHC translation of a contract verification problem for a functional or an imperative
program~\cite{Gr&12,DeAngelisFGHPP21} produces three sets of clauses, as shown
in Figure~\ref{fig:RevCHCs},
where we refer to a program that reverses a list of integers
(we omit the source functional program for lack of space).
The first set (clauses~\mbox{\tts{1}--\tts{4})} is the translation of
the operational semantics of the program.
The second set (clauses \tts{5}--\tts{12}) is the translation
of the properties needed for specifying the contracts.
The third set (goals \tts{13}--\tts{14}) is the translation of the
contracts for the \tts{rev} and \tts{snoc} functions.
In particular, goal \tts{13} is the translation of the contract for
\tts{rev}, which, written in relational form, is the following universally quantified implication:

\tts{$\forall$\,L,R. is\_asorted(L,true) $\wedge$ rev(L,R)  $\rightarrow$ is\_dsorted(R,true)}

\noindent
The atoms \tts{is\_asorted(L,true)} and \tts{is\_dsorted(R,true)} are the precondition and the
postcondition for \tts{rev}, respectively,
stating that, if a list~\tts{L} of integers is sorted in ascending order
with respect to the `$\leq$' relation, then the list \tts{R}
computed by the \tts{rev} function for input \tts{L} is sorted in descending order.

The problem of checking the validity of the contracts for the functions
\tts{rev} and \tts{snoc} is reduced 
to the problem of proving the satisfiability of the set
\ttsi{Reverse} of clauses shown in Figure~\ref{fig:RevCHCs}.
The set \ttsi{Reverse} is indeed satisfiable, but state-of-the-art CHC solvers,
such as Eldarica and SPACER, fail to prove its satisfiability.
This is basically due to the fact that those solvers
lack any form of inductive reasoning on lists and, moreover, they
do not use the information about the validity of the contract for \tts{snoc}  during the proof
of satisfiability of the goal representing the contract for \tts{rev}.

\begin{figure}[ht!]
\begin{sizepar}{8}{10}
\begin{verbatim}
                 /* ------ Program Reverse ------ */
 1. rev([],[]).
 2. rev([H|T],R) :- rev(T,S), snoc(S,H,R).
 3. snoc([],X,[X]).
 4. snoc([X|Xs],Y,[X|Zs]) :- snoc(Xs,Y,Zs).
                 /* ------ Program properties ------ */
 5. is_asorted([],Res) :- Res.
 6. is_asorted([H|T],Res) :- Res = (IsDefHdT => (H=<HdT & ResT)),
                             hd(T,IsDefHdT,HdT), is_asorted(T,ResT).
 7. is_dsorted([],Res) :- Res.
 8. is_dsorted([H|T],Res) :- Res = (IsDefHdT => (H>=HdT & ResT)),
                             hd(T,IsDefHdT,HdT), is_dsorted(T,ResT).
 9. hd([],IsDefHd,Hd) :- ~IsDefHd & Hd=0.    /* hd computes the head of a list.         */
10. hd([H|T],IsDefHd,Hd) :- IsDefHd & Hd=H.  /* IsDefHd=true iff the list is not empty. */
11. leq_all(X,[],Res) :- Res.            /* leq_all(X,L,true) iff for all Y in L, X=<Y. */
12. leq_all(X,[H|T],Res) :- Res = (X=<H & R), leq_all(X,T,R).
                 /* ------ Contracts in goal form ------ */
13. false :- (BL & ~BR), rev(L,R), is_asorted(L,BL), is_dsorted(R,BR). 
14. false :- (BA & BX & ~BC), snoc(A,X,C), is_dsorted(A,BA), 
             leq_all(X,A,BX), is_dsorted(C,BC).
\end{verbatim}
\end{sizepar}
\vspace{-3mm}
\caption{The set {\ttsi{Reverse} of CHCs}. For technical reasons (see Definition~\ref{def:cata}) all program properties are defined by \emph{total functions}. In particular, for the empty list, the \tts{hd} function returns the arbitrarily chosen value \tts{0} (which is never used).\label{fig:RevCHCs}}
\vspace{-3mm}
\end{figure}

The algorithm we will present in Section~\ref{sec:Strategy}
transforms the set \ttsi{Reverse} of clauses into a new set {\ttsi{TransfReverse}} of clauses 
(see Figure~\ref{fig:TransfRevCHCs})
without occurrences of list
terms, such that if {\ttsi{TransfReverse}} is satisfiable, so is {\ttsi{Reverse}}.
Since in the set \ttsi{TransfReverse} there are only integer and boolean terms,
no induction rule is needed for proving its satifiability.

\begin{figure}[ht!]
\flushleft
\begin{sizepar}{8}{10}
\begin{verbatim}
T1. new7(A,B,C,D,E,F,G,H,D,I,J) :- A & B=D & C=(K=>((D>=L) & M)) & E & ~F &
             G=0 & H & (J=((I=<D) & N)) & M & ~K & L=0 & N.
T2. new7(A,B,C,D,E,F,G,H,D,I,J) :- A & B=K & C=(L=>((K>=M) & N)) &
             E=((D=<K) & T) & F & G=K & H=(P=>((K>=Q) & R)) & J=((I=<K) & S) & (R & T)=>N,
             new7(L,M,N,D,T,P,Q,R,D,I,S).
T3. new3(A,B,C,D,E,F) :- A & C & ~D & E=0 & F.
T4. new3(A,B,C,D,E,F) :- D & E=G & F=(H=>((G=<I) & J)) & J=>K & (K & L)=>A,
             new3(K,G,L,H,I,J), new7(M,N,A,G,L,T,P,K,G,B,C).
T5. false :- (A & ~B), new3(B,C,D,E,F,A).
\end{verbatim}
\end{sizepar}
\vspace{-1mm}
\caption{The set {\ttsi{TransfReverse} of transformed CHCs}.
The clauses shown here are those derived from clauses~\tts{1--12} of {\ttsi{Reverse}} and
goal~\tts{13}. The clauses derived from goal~\tts{14} are listed in Appendix~\ref{app:Init-clauses}. 
\label{fig:TransfRevCHCs}}
\vspace{-3mm}
\end{figure}

The transformation works by introducing, for each predicate~$p$
representing a program function (in our case, \tts{rev} and \tts{snoc}),
a new predicate symbol \textit{newp} (in our case, \tts{new3} and \tts{new7})
defined in terms of $p$
together with predicates defining program properties used in the contracts
(in our case,
\tts{is\_asorted}, \tts{is\_dsorted}, \tts{hd}, and \tts{leq\_all}).
The arguments of \textit{newp} are the variables of basic sorts 
occurring in the body of its defining clause, 
and hence \textit{newp}
specifies a relation among the values of the catamorphisms that 
are applied to $p$.
For example, the transformation algorithm introduces the following predicate \tts{new3}: 

\vspace{-1mm}
\begin{sizepar}{9}{10}
\begin{verbatim}
	    new3(K,D,J,G,H,I) :- is_asorted(F,I), hd(F,G,H), rev(F,C),
	                         is_dsorted(C,K), leq_all(D,C,J).
\end{verbatim}
\end{sizepar}
\vspace{-2mm}

\noindent
Then by applying the fold/unfold transformation rules, the algorithm 
derives a recursive definition of \textit{newp}. 
During the transformation, the algorithm makes use of the
user-provided 
contracts as lemmas, thus adding 
new constraints that ease the subsequent satisifiability proof.
By construction, the recursive definition of \textit{newp}
is a set of clauses that do {\it not\/} manipulate ADTs.\,Note 
that while the contract specifications are provided by the users,
the introduction of the new predicate definitions, which is the key step in 
our transformation {algorithm} 
is done in a fully automatic way,
as we will show in Section~\ref{sec:Strategy}.

If the predicates defining program properties are
in the class of catamorphisms (formally defined in Section~\ref{sec:Cata}),
then our transformation is guaranteed to terminate. Thus, we eventually get,
as desired, a set of clauses that are defined on basic sorts only,
whose satisfiability can be checked by CHC solvers that handle the $\textit{LIA}\cup\textit{Bool\/}$ theory.
%
In our example, both Eldarica and SPACER are able to show the satisfiability of \ttsi{TransfReverse}.

	\section{Specifying Contracts using Catamorphisms} 
	\label{sec:Cata}

The notion of a catamorphism has been popularized in the field of functional programming~\cite{MeijerFP91} and many generalizations of it 
have been proposed in the literature~\cite{HinzeWG13}.
Catamorphisms have also been considered in the context of many-sorted first order logic with 
recursively defined functions~\cite{SuterDK10,PhamGW16,GovindSG22},
as we do in this paper.


Let \!$f$ be\! a predicate symbol whose $m \plus n$ arguments (for $m,\!n\!\geq\! 0$)\! 
have
sorts\! $\alpha_1,\!\ldots\!,\!$ $\alpha_m,$ $\beta_1,\ldots,\beta_n$, respectively.
We say that $f$ is {\em functional} from $\alpha_1\times\ldots\times\alpha_m$ to $\beta_1\times\ldots\times\beta_n$,
{with respect to a set $P$ of definite clauses},
if $M(P) \models \forall X, Y, Z.\ f(X,Y) \wedge f(X,Z) ~\rightarrow~ Y\!=\!Z$,


\noindent
where $X$ is an $m$-tuple of distinct variables, and
$Y$ and $Z$ are $n$-tuples of distinct variables.
$X$ and $Y$ are said to be tuples of the {\em input}
and {\em output} variables of $f$, respectively.
Predicate~$f$ is said to be {\em total} 
if $M(P) \models \forall X \exists Y.\ f(X,Y)$.
%
%
%
In what follows, a `total, functional predicate' $f$ from~$\alpha$ to~$\beta$ 
will be called a `total function' and denoted by $f \in \mbox{[}\alpha \rightarrow \beta\mbox{]}$
(the set $P$ of clauses that define $f$ will be understood from the context).

\vspace*{-1mm}
\begin{definition}[Catamorphisms]\label{def:cata}
A {\em list catamorphism}, shown in Figure~\ref{fig:cata3}~(A), is a 
total function \tts{h} $\in [\sigma\times$\tts{list}$(\beta)
\rightarrow \varrho]$, where: (i)~$\sigma$, $\beta$,  and 
$\varrho$ are (products of) basic sorts, 
(ii)~\tts{list}$(\beta)$ is the sort of any list of elements each of which
 is of sort $\beta$,
(iii)~\tts{base1} is a total function in $[\sigma\rightarrow \varrho]$,
    and (iv)~\tts{combine1} is a total function in 
$[\sigma\times\beta\times\varrho\rightarrow \varrho]$.
Similarly, a {\em $($binary$)$ tree catamorphism} (or a {\em tree catamorphism}, for short) 
is a total function 
\tts{t}~$\in [\sigma\times$\tts{tree}$(\beta)\rightarrow \varrho]$ defined 
as shown in Figure~\ref{fig:cata3}~(B).
\end{definition}

\vspace{-2.5mm}

\begin{figure}[ht!]
\begin{sizepar}{9}{10}
\begin{center} 
(A)~~ \begin{subfigure}{.35\textwidth}
{
\begin{verbatim}
h(X,[],Res) :- base1(X,Res).
h(X,[H|T],Res) :- 
    h(X,T,R),
    combine1(X,H,R,Res).
\end{verbatim}
\vspace{-3mm}
}
\end{subfigure}
\hspace*{5mm}
(B)~~ \begin{subfigure}{.44\textwidth}
{\vspace{-1mm}
\begin{verbatim}
t(X,leaf,Res) :- base2(X,Res).
t(X,node(L,N,R),Res) :-  
    t(X,L,RL), t(X,R,RR), 
    combine2(X,N,RL,RR,Res).
\end{verbatim}
\vspace{-3mm}
} 
\end{subfigure}
\end{center}
\end{sizepar}
\vspace{-3mm}
\caption{(A) List catamorphism.   (B) Tree catamorphism.\label{fig:cata3}}
\vspace*{-2.5mm}
\end{figure}

The parameter \tts{X} and some atoms in the body of the clauses in Figure~\ref{fig:cata3}
may be absent (see, for instance, the predicate \tts{hd} in Figure~\ref{fig:RevCHCs}).
The definition of catamorphisms we consider here slightly extends the usual first order definitions~\cite{SuterDK10,PhamGW16,GovindSG22}
by allowing the parameter \tts{X}, which gives 
an extra flexibility for specifying contracts. 
Catamorphisms with parameters have also been considered in 
functional programming~\cite{HinzeWG13}.
To see an example, 
the predicate \tts{leq\_all} 
(see Figure~\ref{fig:RevCHCs})
is a catamorphism in [\tts{int} $\times$ \tts{list}(\tts{int}) $\rightarrow$ \tts{bool}],
where: (i) \tts{base1(X,Res)} is the 
function in [\tts{int} $\rightarrow$ \tts{bool}]
defined by the constraint \tts{Res = true} (i.e.,\,it binds the output boolean variable
\tts{Res}\,to\,\tts{true}), 
and\,(ii)\,\tts{combine1(X,H,R,Res)} is the 
function
in [\tts{int}$\times$\tts{int}$\times$\tts{bool}$\rightarrow$\tts{bool}]
defined by the \!$\textit{LIA}\cup\textit{Bool\/}$ constraint \tts{Res\! =\! \!(X=<H \!\&\! R)}.


The schemata presented in Figure~\ref{fig:cata3}
can be extended by adding to the bodies of the
clauses extra {atoms} 
that have the list tail 
or the left and right subtrees as arguments.
These extensions are shown in Figure~\ref{fig:cataExt},
where \tts{base3}, \tts{combine3}, \tts{base4}, and \tts{combine4}
are total functions on basic sorts, and \tts{f} and \tts{g} are defined by 
instances of the same 
schemata~(C) and~(D), respectively. 
Strictly speaking, these schemata are a CHC translation of 
the {\em zygomorphism} recursion 
schemata~\cite{HinzeWG13}, extended with para\-meter \tts{X}.
However, schemata~(C) and~(D)
can be transformed into schemata~(A) and~(B), 
respectively (see Appendix~\ref{app:Catamorphisms}),  
and hence we prefer not to introduce a different terminology 
and we call them simply 
catamorphisms.

\vspace{-1.5mm}

\begin{figure}[ht!]
\vspace*{-1mm}
\begin{sizepar}{9}{9.8}
\begin{center} 
\hspace*{1mm}(C)~~ \begin{subfigure}{.35\textwidth}
{
\begin{verbatim}
h(X,[],Res) :- base3(X,Res).
h(X,[H|T],Res) :-
   h(X,T,R),
   f(X,T,Rf),
   combine3(X,H,R,Rf,Res).
\end{verbatim}
\vspace{-3mm}
}
\end{subfigure}
\hspace*{6mm}
(D)~~ \begin{subfigure}{.45\textwidth}
{\vspace{-1mm}
\begin{verbatim}
t(X,leaf,Res) :- base4(X,Res).
t(X,node(L,N,R),Res) :-
   t(X,L,RL), t(X,R,RR), 
   g(X,L,RLg), g(X,R,RRg),
   combine4(X,N,RL,RR,RLg,RRg,Res).
\end{verbatim}
\vspace{-3mm}
} 
\end{subfigure}
\end{center}
\end{sizepar}
\vspace{-3.5mm}
\caption{(C) Generalized list catamorphism. (D) Generalized tree catamorphism.\label{fig:cataExt}}
\vspace*{-2mm}
\end{figure}
\vspace*{-1.5mm}

\noindent
Examples of list catamorphisms that are instances of the schema of Figure~\ref{fig:cata3}~(A) are the functions 
\tts{is\_asorted} and \tts{is\_dsorted}
shown in Figure~\ref{fig:RevCHCs} of Section~\ref{sec:IntroExample}.
For instance, \tts{is\_asorted} is a catamorphism in [\tts{list}(\tts{int}) $\rightarrow$ \tts{bool}],
where: (i)~\tts{base3} 
is the constant function defined by the constraint \tts{Res = true},
(ii)~the auxiliary function~\tts{f} is 
\tts{hd}\,$\in$\,[\tts{list(int)}\,$\rightarrow$\,\tts{bool}$\times$\tts{int}], and (iii)~\tts{combine3} is the function
in [\tts{bool}$\times$\tts{int}$\times$\tts{int}$\times$\tts{bool} 
$\rightarrow$ \tts{bool}]
defined by the $\textit{LIA}\cup\textit{Bool\/}$ constraint \tts{Res = (IsDefHdT => (H=<HdT \& ResT))}.
Two more examples are given in Figure~\ref{fig:cataEx3}, where
\tts{count} counts 
the occurrences of a given element in a list,
\tts{bstree} checks whether or not a 
tree is a binary search tree  (duplicate keys are not allowed),
\tts{treemax} and \tts{treemin}  compute, respectively, 
the maximum and~the minimum element in a binary tree.

\vspace{-5mm}
\noindent\begin{figure}[h]
\vspace{-1mm}
\begin{sizepar}{9}{9.7}
\begin{verbatim}
   count(X,[],N) :- N = 0.
   count(X,[H|T],N) :- count(X,T,NT), N = ite(X=H,NT+1,NT).
\end{verbatim}
\vspace*{-3mm}
\begin{verbatim}
   bstree(leaf,B) :- B.
   bstree(node(L,N,R),B) :- bstree(L,BL), bstree(R,BR), treemax(L,IsDefL,MaxL),  
      treemin(R,IsDefR,MinR),  (GrtLeft = (IsDefL => N>MaxL)) &
      (LessRight = (IsDefR => N<MinR)) & (B = (BL & BR & GrtLeft & LessRight)).
\end{verbatim}
\end{sizepar} 
\vspace*{-5mm}
\caption{The catamorphisms {\tts{count} and \tts{bstree}}. 
 \label{fig:cataEx3}}
\vspace{-4mm}
\end{figure}

\noindent
In the sets of CHCs we consider, we identify 
two disjoint sets of predicates: 
(1)~the {\em program predicates}, defined by any set of CHCs not containing occurrences of catamorphisms, and (2)~the {\em catamorphisms}, defined by instances of the schemata in Figure~\ref{fig:cataExt}.
An atom is said to be a {\textit{program atom}} (or a {\textit{catamorphism atom}}) if its predicate symbol
is a program predicate (or a catamorphism, respectively).

\label{subsec:Contracts}
\begin{definition}
A {\it{contract}}  \label{def:contract-spec}
 is a formula of the form (where the implication is right-associative): 


\hspace{3mm}$({\mathit{K}})$\hspace{5mm} $\mathit{pred}(Z) \rightarrow c, \mathit{cata}_1(X_1,T_1,Y_1),\ldots,\mathit{cata}_n(X_n,T_n,Y_n) \rightarrow d$ 

\noindent
where: (i)~$\mathit{pred}$ is a program predicate and $\mathit{Z}$ is a tuple of distinct variables,
(ii)\,$c$ is a constraint such that $\mathit{vars}(c)\!\subseteq\! 
\{X\!_{1},\ldots,X\!_{n},Z\}$, 
(iii)~$\mathit{cata}_1,\ldots,$ $\mathit{cata}_n$ are catamorphisms,
(iv)~${X_1},\ldots,{X_n},$ {${Y_1},\ldots,{Y_n}$ are pairwise} disjoint tuples of distinct variables of basic sort,
(v)~${T_1},\ldots,{T_n}$ are ADT variables 
occurring in $Z$, and 
(vi)~$d$ is a constraint, called the {\em postcondition} of the contract, such that $\mathit{vars}(d) \subseteq \{X_1,\ldots,X_n,$ $Y_1,\ldots,Y_n,Z\}$.
\end{definition}

\vspace{-1mm}
The following are the contracts for \tts{rev} and 
\tts{snoc}\footnote{In concrete contract specifications
we use the keyword `\tts{spec}' and the two distinct implication symbols 
`\tts{==>}' and `\tts{=>}'.}.

\vspace{1mm}

\noindent
\begin{sizepar}{9}{10}
\noindent 
\tts{:- spec rev(L,R) ==> is\_asorted(L,BL), is\_dsorted(R,BR) => (BL=>BR).}

\noindent 
\tts{:- spec snoc(A,X,C) ==> is\_dsorted(A,BA), leq\_all(X,A,BX), is\_dsorted(C,BC)}

\noindent 
\hspace{40mm}\tts{=> ((BX \& BA) => BC).} 
\end{sizepar}
\vspace{-2mm}
\begin{definition}\label{def:valid}
Let $\mathit{Catas}$ denote the conjunction
$\mathit{cata}_1(X_1,T_1,Y_1) \wedge \ldots \wedge 
\mathit{cata}_n(X_n,T_n,Y_n)$ of the catamorphisms in the contract~$K$ (see Definition~\ref{def:contract-spec}),
and let~$P$ be a set of definite CHCs.
We say that contract~$K$ is {\em valid} $($with respect to the set $P$ of CHCs$)$ if 
$M(P) \models \forall \ (\mathit{pred}(Z) \wedge c \wedge \mathit{Catas}
~\rightarrow~ d)$.
\end{definition}
%

\vspace*{-3mm}
\begin{theorem}[Correctness of the CHC translation]\label{thm:contract}
For contract $K$, let $\gamma(K)$ denote the goal $\mathit{false} \leftarrow \neg\,d, c, \mathit{pred}(Z), \mathit{Catas}$. Contract~$K$ is valid with respect to a set $P$ of CHCs 
if and only if 
$P\cup \{\gamma(K)\}$ 
is satisfiable.
\end{theorem}


The proof of this theorem is given in Appendix~\ref{app:Proofs}.  
The use of the catamorphism schemata (C) and (D) guarantees the termination of the transformation algorithm (see Theorem~\ref{thm:termination})
and, at same time, allows the specification of many nontrivial contracts
(see our benchmark in Section~\ref{sec:Experiments}). 
Among the properties that cannot be specified by our notion of catamorphisms,
we mention ADT equality (indeed ADT equality 
has {\em more than one} ADT argument).
Thus, in particular, the property \tts{$\forall$\,L,RR. double-rev(L,RR) 
$\!\rightarrow\!$ L=RR}, where \tts{double-rev(L,RR)} holds if the 
conjunction `\tts{rev(L,R),\,rev(R,RR)}' holds, cannot be written as a 
contract in our framework. 
We leave it for future work to identify larger classes of
 contracts that can be handled by our transformation-based approach.

\vspace*{-4mm}



%
%
%
%
%

	\section{Catamorphism-based Transformation Algorithm} 
	\label{sec:Strategy}
	In this section we present Algorithm \Cata~(see Figure~\ref{fig:AlgoC})  
which, 
given a set~$P$ of definite clauses manipulating ADTs and a set ${\mathit{Cns}}$ of contracts,
derives 
a set $\mathit{TransfCls}$ of CHCs 
with new predicates manipulating terms of basic sorts only, such that
if $\mathit{TransfCls}$ is satisfiable, then $P\cup \{\gamma(K) \mid K\!\in\! \mathit{Cns}\}$
is satisfiable.
By Theorem~\ref{thm:contract},  
the satisfiability of $\mathit{TransfCls}$ implies
the validity of the contracts in ${\mathit{Cns}}$  with respect to~$P$.

\smallskip

%


\begin{figure}[!ht]
%
%
\noindent \hrulefill\nopagebreak

\vspace{2mm}

\begin{minipage}{130mm}
\noindent {\bf Algorithm}~\Cata.\\
{\em Input}: A set $\mathit{P}$ of definite clauses and a set $\mathit{Cns}$ of contracts, one for each program predicate.

{\em Output}: A set $\mathit{TransfCls}$ of clauses (including goals) 
on basic sorts such that, if $\mathit{TransfCls}$ is satisfiable, then 
every contract in $\mathit{Cns}$ is valid with respect to $P$.

\vspace*{-2mm}
\noindent \rule{2.0cm}{0.2mm}

\noindent 

\noindent $\mathit{InCls}:=\{\gamma(K) \mid K\!\in\! \mathit{Cns}\}$;~~ $\mathit{Defs}:=\emptyset$;~~
\noindent $\mathit{OutCls}:=\mathit{InCls};$

\noindent
{\bf while}~ $\mathit{InCls}\!\neq\!\emptyset$ ~{\bf do}~~
\begin{minipage}[t]{99mm} 
\hspace{0mm}\makebox[50mm][l]{$\mathit{Define}(\mathit{InCls},\mathit{Defs},\mathit{NewDefs});$}

\hspace{0mm}$\mathit{Unfold}(\mathit{NewDefs},\mathit{P},\mathit{UnfCls});$

\hspace{0mm}$\mathit{Apply}\mbox{-}\mathit{Contracts}(\mathit{UnfCls}, \mathit{Cns}, \mathit{RCls});$

\hspace{0mm}\makebox[58mm][l]{$\mathit{InCls}:=\mathit{not\mbox{-}foldable(RCls,Defs)};$}

\hspace{0mm}$\mathit{OutCls}:= \mathit{OutCls}\cup\mathit{foldable(RCls,Defs)};$
%
%
\end{minipage} 
$\mathit{Fold}(\mathit{OutCls},\mathit{Defs}, \mathit{TransfCls})$

\vspace*{-2mm} 
\noindent \hrulefill
\vspace*{-2mm} 
\caption{The Transformation Algorithm~\Cata \label{fig:AlgoC}.}
\end{minipage}
\vspace*{-2mm}
\end{figure}

During the {\bf while-do} loop, \Cata~iterates the
$\mathit{Define\mbox}$, $\mathit{Unfold}$, and 
$\mathit{Apply}$-$\mathit{Contracts}$ procedures 
as we now explain.
Their formal definition is given in 
Figures~\ref{fig:Define}, \ref{fig:unfoldProc}, and \ref{fig:contractProc}.

\smallskip
\noindent
-~Procedure $\mathit{Define}$ 
works by introducing suitable new predicates defined by clauses, called {\em definitions},
of the form: $\mathit{newp}(U) \If c, A, \mathit{Catas}\!_{A}$,
where~$U$ is a tuple of variables of basic sort, $A$ is a program atom,
and $\mathit{Catas}\!_A$ is a conjunction of catamorphism atoms whose ADT variables
occur in $A$. 
Thus, $\mathit{newp}(U)$ defines the projection onto $\textit{LIA}\cup\textit{Bool\/}$
of the relation between the variables of the program atom $A$
and the catamorphisms acting on the ADT variables of~$A$.
In particular, 
for each program atom $A$ occurring in the body `$c,G$\hspace*{1pt}' 
of a definite clause or a goal in $\mathit{InCls}$,
the $\mathit{Define}$ procedure may {\em either} 
(i)~introduce a new definition 
whose body consists of~$A$, 
together with the conjunction of {all} catamorphism atoms in $G$
that share an ADT variable with $A$, and the constraints on the input 
variables of $A$  {of basic sort}  (case {\it Project\/})
{\em or} (ii)~extend an already introduced definition 
for the program predicate of~$A$
by (ii.1)~adding new catamorphism atoms to its body 
and/or (ii.2) generalizing its constraint 
(case {\it Extend\/}).
The new predicate definitions introduced by\,a single 
application of the {\it $\mathit{Define\mbox}$} procedure 
are collected in $\mathit{NewDefs}$,
while the set of all definitions introduced during the 
execution of~\Cata~are collected in~$\mathit{Defs}$.


\smallskip
\noindent
-~Then, Procedure $\mathit{Unfold}$ 
 (1) unfolds the program atoms occurring in the body of 
the definitions belonging to $\mathit{NewDefs}$,  and then 
(2) unfolds all catamorphism atoms whose ADT argument is not a variable.
Finally, (3) by the functionality property 
(see Section~\ref{sec:Cata}), repeated occurrences of a function with the same input
are removed.


\smallskip
\noindent
-~Next, Procedure 
$\mathit{Apply}$-$\mathit{Contracts}$ 
applies the contracts  in
${\mathit{Cns}}$ for the program predicates occurring in the body of the clauses 
generated by the $\mathit{Unfold}$ procedure.
This is done in two steps. First, (1) for each program atom $A$ 
in a clause~$C$ obtained by unfolding, the procedure adds 
the catamorphism atoms (with free output variables) that are present
in the contract for 
$A$ and not in the body of $C$. 
Note that, since catamorphisms are {\it total} functions,
their addition preserves satisfiability of clauses.
Then, (2) if the constraints on the input variables of the added catamorphisms are satisfiable,
the procedure adds also the postconditions of the contracts.
Thus, the effect of the $\mathit{Apply}$-$\mathit{Contracts}$ procedure 
is similar to the one produced by the application of user-provided 
lemmas.

\smallskip

The $\mathit{Unfold}$ and $\mathit{Apply}$-$\mathit{Contracts}$ procedures
may generate clauses that are not \mbox{{\em foldable},} that is, clauses 
whose body `$c,G$\hspace*{1pt}' contains a conjunction of a program atom and catamorphism atoms
which is not a variant of the body of any definition in {\it Defs},
or, if there is such a definition in {\it Defs}, the constraint $c$ does not 
imply the constraint occurring in that definition.
By using the function $\mathit{not\mbox{-}foldable},$ those clauses
are added to the set $\mathit{InCls}$ of clauses to be further processed
by the {\bf while-do} loop,
while the others, by using the function~$\mathit{foldable}$, 
are added to the set $\mathit{OutCls}$ of clauses 
that are
output by the 
loop. 


%

\smallskip

The termination of the {\bf while}-{\bf do} loop is guaranteed by the following two facts:
(i)~there are finitely many catamorphism atoms that can be added to the body of a definition, and 
(ii) by implementing constraint generalization through a {\em widening} operator~\cite{CoH78},
a most general constraint is eventually computed. %
When \mbox{Algorithm~\Cata} exits the {\bf while}-{\bf do} loop, it
returns a set $\mathit{OutCls}$ of clauses (which are all foldable) 
and a set $\mathit{Defs}$ of new predicate definitions.
Then, the  $\mathit{Fold}$ procedure (\mbox{Figure~\ref{fig:foldProc})}
uses the definitions in $\mathit{Defs}$
for removing ADT variables from the clauses in $\mathit{OutCls}$.
By construction, (i) the head of each clause~$C$ in $\mathit{OutCls}$
is either {\it false} or an atom $\mathit{newq}(V)$, where $V$ is a tuple of variables of basic sort, 
and (ii) for each conjunction of a program atom $A$ and catamorphism atoms $B$ 
in the body of $C$ {sharing an ADT variable with A},
there is in $\mathit{Defs}$ a definition $\mathit{newp}(U) \If c, A, 
\mathit{Catas}\!_A$ 
such that $B$ is a subconjunction of $\mathit{Catas}\!_A$ and $c$ is implied by
the constraint in~$C$. The {\it Fold} procedure removes all ADT 
variables by replacing in $C$ the atom~$A$ by $\mathit{newp}(U)$
and removing the subconjunction $B$. 

\smallskip

Let us introduce some notions used in the procedures {\it Define}, {\it Unfold},
$\mathit{Apply}$-$\mathit{Contracts}$, and {\it Fold}.
Given a conjunction $G$ of atoms, by $\mathit{bvars}(G)$ and 
$\mathit{adt\mbox{-}vars}(G)$
we denote the set of variables in~$G$ that have a basic sort and an ADT sort, respectively.

\begin{definition}
The {\em projection} of a constraint
	$c$ onto a tuple $V$ of variables is a constraint $\pi(c,V)$ such that: (i)~$\mathit{vars}(\pi(c,V))\!\subseteq\! V$ and
	(ii)~$\mathbb D \models \forall (c\!\rightarrow\!\pi(c,V))$. 
	
A {\em generalization}  of two constraints $c_1$ and $c_2$ is a constraint, denoted $\alpha (c_1,c_2)$,
such that $\mathbb D \models\forall (c_1 \!\rightarrow \alpha (c_1,c_2))$ and 
$\mathbb D \models\forall (c_2 \!\rightarrow \alpha (c_1,c_2))$.

Let $D$$:$ $\mathit{newp}(U) \If c, {A}, \mathit{Catas}\!_A$ be a clause in {\it Defs}, 
where: (i)~$A$ is a program atom with predicate $p$,
(ii)~$\mathit{Catas}\!_A$ is a conjunction of catamorphism atoms, and
(iii)~$c$ is a constraint on input variables {of $A$}, 
and $U$ is a tuple of
variables of basic sort.
We say that $D$ is {\em maximal} {for $p$} 
if, for all defin\-i\-tions 
$\mathit{newq}(V) \If d,{A}, B$ in $\mathit{Defs}$, we have that:
(i)~$B$ is a subconjunction of  $\mathit{Catas}\!_A$,
(ii)~\mbox{$\mathbb D\!\models\!\forall (d\! \rightarrow\! c)$}, and
(iii)~$V$ is a subtuple of $U$.
\end{definition}



{For a concrete definition of a generalization operator, based on widening, 
we refer to the existing literature~\cite{DeAngelisFGHPP21}. }

Note that, by the {\it Extend\/} case of the {\it Define} procedure, for every program 
predicate $p$ occurring in $\mathit{Defs}$,
there is a unique maximal definition.

\vspace*{-2mm}
\begin{figure}[!ht]
\begin{minipage}{135mm}
\noindent \hrulefill \nopagebreak

\noindent {\bf Procedure $\mathit{Define}(\mathit{InCls},\mathit{Defs}, 
\mathit{NewDefs})$}
\\
{\em Input}\/: A set $\mathit{InCls}$ of clauses and a set $\mathit{Defs}$ of definitions.
\\
{\em Output}\/: A set $\mathit{NewDefs}$ of new definitions.

\vspace{-2mm}
\noindent \rule{2.0cm}{0.2mm}

\noindent $\mathit{NewDefs} := \emptyset$; 

\noindent {\bf for} each clause $C$: $H\leftarrow c, G$ in $\mathit{InCls}$ 
and each program atom $A$ in $G$ {\bf do}

\smallskip

\hspace{2mm}
\begin{minipage}{132mm}
\noindent
let $\mathit{Catas}\!_A=\bigwedge \{F \mid F\mbox{ is a catamorphism atom in } G \mbox{ and } \mathit{adt\mbox{-}vars}(A) \cap 
\mathit{adt\mbox{-}vars}(F) \neq \emptyset\}$;

\smallskip

({\it Clause is foldable}) {\bf if} in $\mathit{Defs}$ there is a clause $\mathit{newp}(U) \If d, 
    A, B$, 
     with  $U\!=\!\mathit{bvars}(\{d,A,B\})$,
such that:
(i)~$\mathit{Catas}\!_A$ is a subconjunction of $B$,
and (ii)~$\mathbb D \models\forall (c \rightarrow d)$, 
{\bf then}  skip;

\smallskip

({\it Extend}) {\bf else if} the maximal definition for the predicate of $A$  in $\mathit{Defs}\cup \mathit{NewDefs}$
is a clause $\mathit{newp}(U) \If d, A, B$,
such that: 
(i) $\mathit{Catas}\!_A$ is {\em not} a subconjunction of $B$, or
(ii)\,$\mathbb D \not\models\forall (c \rightarrow d)$,
{\bf then}
{introduce definition} $\mathit{ExtD}$: $\mathit{extp}(V) \If \alpha(d,c), A, B'$,
where: (i) $\mathit{extp}$ is a new predicate symbol,
(ii) $V\!=\!\mathit{bvars}(\{\alpha(d,c),A,B'\})$, and
$B'\!=\!\bigwedge\{F\mid F$ occurs either in $B$ or in $\mathit{Catas}\!_A\}$; 
 ~~ 
$\mathit{NewDefs} := \mathit{NewDefs} \cup \{\mathit{ExtD}\}$;~~~ $\mathit{Defs} := \mathit{Defs} \cup \{\mathit{ExtD}\}$;

\smallskip

({\it Project}) {\bf else if} there is no clause in $\mathit{Defs}$ with $A$ in its body

{\bf then} {introduce definition} $D$: $\mathit{newp}(U) \If \pi(c,I), A, \mathit{Catas}\!_A$,
where: 
(i)~$\mathit{newp}$ is a new predicate symbol,
(ii)~$I$ is the tuple of the input variables of basic sort in $\{A, \mathit{Catas}\!_A\}$, and
(iii)~$U\!=\!\mathit{bvars}(\{\pi(c,I), A, \mathit{Catas}\!_A\})$;~~~
$\mathit{NewDefs} := \mathit{NewDefs} \cup \{D\}$;~~~ $\mathit{Defs} := \mathit{Defs} \cup \{D\}$;

\smallskip

\end{minipage}


\noindent \hrulefill
\vspace*{-2mm}
\caption{The $\mathit{Define}$ procedure.
\label{fig:Define}}
\end{minipage}
\vspace*{-5mm}
\end{figure}

\begin{example}[Reverse, continued]\label{ex:rev2}
$\mathit{InCls}$ is initialized to the set \{\tts{13}, \tts{14}\} of goals (see
Figure~\ref{fig:RevCHCs}). 
The {\bf while-do} loop starts by applying the $\mathit{Define}$ procedure to
goal~\tts{13}. (For lack of space, we will not show here the transformation steps starting from goal \tts{14}.)
No definitions for predicate \tts{rev} are present in
$\mathit{Defs}$, and hence the case {\it Project\/} applies. Thus,
the $\mathit{Define}$ procedure introduces the 
following clause defining the new predicate \tts{new1}:

\vspace{-1mm}
\begin{sizepar}{9}{10} 
\begin{verbatim}
D1. new1(BL,BR) :- is_asorted(L,BL), rev(L,R), is_dsorted(R,BR).
\end{verbatim}
\end{sizepar} 

\vspace{-1.5mm}

\noindent
where: (i)~the body is made out of the program atom \tts{rev(L,R)} 
and the catamorphisms on the lists~\tts{L} and \tts{R} occurring in goal~\tts{13},
(ii)~\tts{BL} and \tts{BR} are the variables of basic sort in the body of \tts{D1},  and
(iii)~the projection of the constraint of goal \tts{13} onto the (empty) tuple of input variables of basic sort of the body of \tts{D1} is 
\tts{true} (and thus, omitted). \hfill $\Box$. 
\end{example}


%




\vspace{-2.5mm}
\begin{definition}[Unfolding]
Let  $C$: $H\leftarrow c,G_L,A,G_R$ be a clause, where $A$ is an atom,
and let $P$ be a set of definite clauses with 
$\mathit{vars}(C)\cap\mathit{vars}(P)=\emptyset$.
Let {\it Cls}: $\{K_{1}\leftarrow c_{1},
B_{1},~\ldots,~K_{m}\leftarrow c_{m}, B_{m}\}$, with $m\!\geq\!0$,
be the set of clauses in $P$,
such that: for $j=1,\ldots,m$,
$(i)$~there exists a most general unifier~$\vartheta_j$ of $A$ 
and~$K_j$, and {$(ii)$~the conjunction of constraints $(c, c_{j})\vartheta_j$ is satisfiable.}
We define: $\mathit{Unf}(C,A,P)=\{(H\leftarrow  c, {c}_j,G_L, B_j, G_R) 
\vartheta_j \mid  j=1, \ldots, m\}.$
\end{definition}

\vspace*{-3mm}
\begin{example}[Reverse, continued]\label{ex:rev-continued} \nopagebreak
	The ${\mathit{Unfold}}$ procedure first (1) unfolds
	the program atom \tts{rev(L,R)} in clause~\tts{D1}, and then
	(2) unfolds the catamorphism atoms with non-variable  ADT arguments. We get:
	
	\vspace{-2mm}
	\begin{sizepar}{9}{11} 
		\begin{verbatim}
		C1.  new1(A,B) :- A & B.
		C2.  new1(A,B) :- A=(G=>((D=<H) & I)), 
		       is_asorted(F,I), hd(F,G,H), rev(F,C), snoc(C,D,E), is_dsorted(E,B).
		\end{verbatim}
	\end{sizepar} 
	\vspace{-1.5mm}
	\noindent
	Finally, functionality is not applicable, and $\mathit{Unfold}$ terminates.
	\hfill $\Box$
\end{example}

\vspace{-1mm}
\begin{figure}[!ht]
\begin{minipage}{135mm}
\noindent \hrulefill

\noindent {\bf Procedure $\mathit{Unfold}(\mathit{NewDefs},P,\mathit{UnfCls})$}
\\
{\em Input}\/: A set $\mathit{NewDefs}$ of definitions and a set $P$ of definite clauses.
\\
{\em Output}\/: A set $\mathit{UnfCls}$ of clauses.

\vspace*{-2.5mm} \noindent \rule{2.0cm}{0.2mm}

\noindent
$\mathit{UnfCls} := \mathit{NewDefs}$; 

\noindent
(1)\hspace{1mm}(\!{\it{Unfold program atom}}) 
\ {\bf for} all clauses $D$: $\mathit{newp}(U) \If c, A, \mathit{Catas}\!_A$ in $\mathit{UnfCls}$, 
where $A$ has a program predicate {\bf do} $\mathit{UnfCls}:=(\mathit{UnfCls}-\{D\}) \cup \mathit{Unf(D,A,P)}$;

\noindent
(2)\hspace{1mm}(\!{\it{Unfold catamorphisms}})
\ {\bf while} there exists a clause $C$: \mbox{$H\! \leftarrow d, {L}, B, {R}$} in $\mathit{UnfCls}$, for some conjunctions\,$L$ and~$R$ of atoms,
such that $B$ is a catamorphism atom whose input argument of ADT sort is not a variable {\bf do}
$\mathit{UnfCls}:=(\mathit{UnfCls}-\{C\}) \cup \mathit{Unf(C,B,P)}$;~

\noindent
(3)\hspace{1mm}(\!{\it{Apply Functionality}}) 
\ {\bf while} there exists a clause $C$:\,\mbox{$H\!\leftarrow d, {L}, h(X,Y), h(X,Z),{R}$} 
 in $\mathit{UnfCls}$,  for some catamorphism $h$ {\bf do}
\
$\mathit{UnfCls}:=(\mathit{UnfCls}-\{C\}) \cup \{H\! \leftarrow d, Y\!\!=\!Z, {L}, h(X,Y), {R}\}$; 
\end{minipage}

\noindent \rule{0mm}{1mm}\hrulefill
\vspace{-2.5mm}
\caption{The ${\mathit{Unfold}}$ procedure.\label{fig:unfoldProc}}
\vspace{-1mm}
\end{figure}


\vspace{-2.mm}
\begin{figure}[!ht]
\begin{minipage}{135mm}
\noindent \hrulefill

\noindent {\bf Procedure {\it  Apply-Contracts}$(\mathit{UnfCls}, \mathit{Cns}, \mathit{RCls})$}
\\
{\em Input}\/: A set $\mathit{UnfCls}$ of clauses and a set $\mathit{Cns}$ of contracts, 
one for each program predicate.
\\
{\em Output}\/: A set $\mathit{RCls}$ of clauses.

\vspace*{-2.5mm} \noindent \rule{2.0cm}{0.2mm}

$\mathit{RCls} := \emptyset$;

\noindent {\bf for} each clause $C$: $H\leftarrow e, G$ in $\mathit{UnfCls}$ {\bf do}

\noindent 
\hspace{3mm}(1) (\!{\it{Catamorphism Addition}})
\ {\bf for} each program atom $A$ in $G$ {\bf do}

\smallskip

\hspace{2mm}
\begin{minipage}{132mm}
\noindent

\hangindent=3mm
\makebox[3mm]{-}let ${\mathit{K}}$: $A\rightarrow c, \mathit{cata}_1(X_1,T_1,Y_1),\ldots,
\mathit{cata}_n(X_n,T_n,Y_n) \rightarrow d$ be the contract in  $\mathit{Cns}$
for the predicate of $A$, with $Y_1,\ldots,Y_n$ 
variables not occurring in $C$; 

\hangindent=3mm
\makebox[3mm]{-}{\bf for} each catamorphism atom $\mathit{cata}_i(X_i,T_i,Y_i)$ in ${\mathit{K}}$ {\bf do}

~~~~~~{\bf if} there is no atom $\mathit{cata}_i(V,T_i,W)$ in  $G$ 
{\bf then} add $\mathit{cata}_i(X_i,T_i,Y_i)$ to $G$; 

\end{minipage}

\smallskip

\noindent 
\hspace{3mm}(2) {(\!{\it{Constraint Addition}})}
\ let $E$: $H\leftarrow e, G'$ be the clause obtained at the end of Step~(1);

\noindent 
\hspace{3mm}\begin{minipage}{132mm}

\smallskip

- let $c_1,\ldots,c_k$ be the {constraints on the input variables of the catamorphisms}
in the contracts \hangindent=2mm
used for deriving $E$, and 
let $d_1,\ldots,d_k$ be the corresponding contract
postconditions;

\hangindent=2mm
- let $Z$ be the tuple of variables occurring in $\{c_1,\ldots,c_k\}$ and not in $E$;

\hspace*{2mm}\hangindent=2mm
{\bf if} $\mathbb{D}\models \forall(e \rightarrow \exists Z.\, c_1\wedge \ldots \wedge c_k)$ 
{\bf then} 
$\mathit{RCls} := \mathit{RCls} \cup \{H\leftarrow e, c_1,\ldots,c_k,d_1,\ldots,d_k, G'\}$
\smallskip
\end{minipage}
\end{minipage}

\vspace*{-.5mm}
\noindent \hrulefill
\vspace{-3mm}
\caption{The $\mathit{Apply\mbox{-}Contracts}$ procedure.\label{fig:contractProc}}
\vspace*{-2mm}
\end{figure}

\vspace*{-1mm}
\begin{example}[Reverse, continued]\label{ex:rev-continued2} 
The {\it  Apply-Contracts} procedure first (1) adds the catamorphism atoms 
\tts{is\_dsorted(C,K), leq\_all(D,C,J)} to the body of clause \tts{C2}, and
then (2) adds the postconditions of
the contracts for \tts{rev} and \tts{snoc}. We get:
\vspace{-1mm}

\begin{sizepar}{9}{11} 
\begin{verbatim}
C3. new1(A,B) :- A=(G=>(D=<H & I)) & I=>K & (K&J)=>B, is_asorted(F,I), hd(F,G,H), 
       rev(F,C), snoc(C,D,E), is_dsorted(C,K), leq_all(D,C,J), is_dsorted(E,B).
\end{verbatim}
\end{sizepar}  
\vspace{-1.5mm}

\noindent
Now, in the second iteration of the
{\bf while}-{\bf do} loop of \Cata, 
clause \tts{C3} is processed by the $\mathit{Define}$
procedure. The following new definition \tts{D2} relative to the program atom \tts{rev(F,C)}, is introduced according to the
 {\it Extend\/} case
because the catamorphism atoms 
in 
\tts{C3} that share ADT variables with \tts{rev(F,C)}
is not a subset of the catamorphism atoms in 
\tts{D1}.

\vspace{-1mm}
\begin{sizepar}{9}{10} 
\begin{verbatim}
D2. new3(K,D,J,G,H,I) :- is_asorted(F,I), hd(F,G,H), rev(F,C), 
        is_dsorted(C,K), leq_all(D,C,J).
\end{verbatim}
\end{sizepar} 

\vspace*{-7mm}\hfill$\Box$
\vspace{-1mm}
\end{example}

\begin{figure}[!ht]
\begin{minipage}{135mm}
\noindent \hrulefill

\noindent {\bf Procedure $\mathit{Fold}(\mathit{OutCls},\mathit{Defs},\mathit{TransfCls})$}
\\
{\em Input}\/: A set $\mathit{OutCls}$ of clauses and a set $\mathit{Defs}$ of definitions. 
\\
{\em Output}\/: A set $\mathit{TransfCls}$ of clauses.

\vspace*{-2.5mm} \noindent \rule{2.0cm}{0.2mm}

$\mathit{MDefs} := \{D \! \in\! \mathit{Defs} \mid D$ is a maximal definition for some program predicate$\}$;

\hangindent=2.6mm
$\mathit{TbF} := \{C\! \in\! \mathit{OutCls} \mid$ the head of $C$ is either 
$\mathit{false}$  or its predicate occurs in {a head of} {\it MDefs}$\}$;

\noindent 
{\bf for} each clause $C$: $H\leftarrow e, G$ in $\mathit{TbF}$ and  each program atom $A$ in $G$ {\bf do}


\hspace{2mm}
\begin{minipage}{132mm}    
\noindent
	
\hangindent=3mm
\makebox[3mm]{-}let $B\!_A=\bigwedge \{F \mid F\mbox{ is a catamorphism atom in } G \mbox{ and } \mathit{adt\mbox{-}vars}(A) \cap 
\mathit{adt\mbox{-}vars}(F) \neq \emptyset\}$;

\hangindent=3mm
\makebox[3mm]{-}let $D$: $\mathit{newp}(U) \If c, A, \mathit{Catas}\!_A$  be the definition in $\mathit{MDefs}$ 
for the predicate of~$A$, such that $\mathbb{D}\models \forall (e\rightarrow c)$ 
and $B_{A}$ is a subconjunction of $\mathit{Catas}\!_A$;

\hangindent=3mm
\makebox[3mm]{-}replace $A$ by $\mathit{newp}(U)$ in the body of $C$;
\end{minipage}

\vspace*{.3mm}
\noindent 
Remove all catamorphism atoms from the derived clauses
and add them to $\mathit{TransfCls}$;
\end{minipage}
	
	\vspace*{0.5mm}
	\noindent \rule{0mm}{1mm}\hrulefill   
	\vspace{-2mm}
	\caption{The $\mathit{Fold}$ procedure.\label{fig:foldProc} }
	\vspace*{-3mm}
\end{figure}

\vspace{-2mm}
\begin{example}[Reverse, continued]\label{ex:rev3} 
When Algorithm \Cata~exits the {\bf while}-{\bf do} loop, we get a set $\mathit{OutCls}$ of clauses including:


\vspace{-1mm}
\begin{sizepar}{9}{11} 
\begin{verbatim}
C4. new3(A,B,C,D,E,F) :- (D & E=G & F=(H=>G=<I & J) & J=>K & (K&L)=>A),
       is_asorted(M,J), hd(M,H,I), rev(M,N), is_dsorted(N,K), leq_all(G,N,L),
       snoc(N,G,V), is_dsorted(V,A), leq_all(B,V,C).
\end{verbatim}
\end{sizepar} 

\vspace{-1.5mm}

\noindent
and a set $\mathit{Defs}$ of definitions including clause \tts{D2} and the following clause 
relative to \tts{snoc}:


\vspace{-1.5mm}
\begin{sizepar}{9}{11} 
\begin{verbatim}
D3. new7(A,B,C,D,E,F,G,H,D,I,J) :- hd(K,A,B), is_dsorted(K,C), leq_all(D,L,E), 
       hd(L,F,G), is_dsorted(L,H), snoc(L,D,K), leq_all(I,K,J).
\end{verbatim}
\end{sizepar} 
\vspace*{-2mm}

\noindent
By the {\it Fold} procedure, from clause \tts{C4}, using 
\tts{D2} and \tts{D3}, we get clause \tts{T4} of Figure~\ref{fig:TransfRevCHCs}.  \hfill$\Box$
\end{example}


\vspace{-1mm}
\begin{theorem}[Termination of {Algorithm}~\Cata] 
\label{thm:termination}
Let $\mathit{P}$ be a set of definite clauses and $\mathit{Cns}$ a set of contracts
specified by catamorphisms.
Then, Algorithm~\Cata~terminates for $\mathit{P}$ and $\mathit{Cns}$.
\end{theorem}

\vspace{1mm}

\begin{theorem}[Soundness of {Algorithm}~\Cata] 
\label{thm:soundness-AlgorithmR}
Let $\mathit{P}$ and $\mathit{Cns}$ be the input of {Algorithm}~\Cata,
and let $\mathit{TransfCls}$ be the output set of clauses.
Then, every clause in $\mathit{TransfCls}$ has basic sort, and
if $\mathit{TransfCls}$ is satisfiable, then all contracts 
in $\mathit{Cns}$ are valid with respect to $P$.
\end{theorem}

\vspace{-1.5mm}

The proofs of Theorems 2 and 3 are given in Appendix~\ref{app:Proofs}. 
The converse of Theorem~\ref{thm:soundness-AlgorithmR} does not
hold.
Thus, the unsatisfiability of $\mathit{TransfCls}$ 
means
that our transformation technique is unable 
to prove the validity of the contract at hand,
and not necessarily that the contract is not~valid.
\vspace*{-5.5mm}

	\section{Experimental Evaluation} 
	\label{sec:Experiments}
\newcommand{\vericat}{{VeriCa\hspace{-0.3mm}T}\xspace}

In this section we describe a tool, called \vericat, implementing our
verification method based on the use of catamorphisms
and we present some 
case studies to which \vericat has been successfully applied
(see 
\tts{\url{https://fmlab.unich.it/vericat/}}
for details).

\vericat implements the two {steps} of our method:
(i)~the \textit{Transf} {step}, 
which realizes the CHC transformation algorithm
presented in Section~\ref{sec:Strategy}, and (ii)~the \textit{CheckSat} step,
which checks CHC satisfiability.
For the \textit{Transf} {step},
\mbox{\vericat} uses VeriMAP~\cite{De&14b}, a tool for
fold/unfold transformation of CHCs.
{It takes as input a set of CHCs 
representing a program manipulating ADTs, together with its contracts, and
returns a new set of CHCs acting on variables of basic sorts only.}
For the \textit{CheckSat} {step}, \mbox{\vericat} uses SPACER to check the
satisfiability of the transformed CHCs.


We have applied our method to prove the validity of contracts for programs
{implementing various algorithms}
for concatenating, permuting, reversing, and sorting lists of integers,
and also for inserting and deleting elements in binary search trees.
The contracts specify properties defined by catamorphisms such as:
list length, tree size, tree height, list (or tree) minimum and maximum element,
list sortedness (in ascending or descending order), {binary search tree property,}
element sum, and list (or tree) content (defined as sets or
multisets of elements).
For instance, for the sorting programs implementing
Bubblesort, Insertionsort, Mergesort, Quicksort, Selectionsort, and Treesort,
\vericat was able to prove the following two contracts:

\vspace{-.5mm}
\noindent
~~~~~\tts{:- spec sort(Xs,Ys) ==> is\_asorted(Ys,B) => B.}

\vspace{-.5mm}
\noindent
stating that the output \tts{Ys} of \tts{sort} is a list in ascending order,
and

\vspace{-.5mm}
\noindent
~~~~~\tts{:- spec sort(Xs,Ys) ==> count(X,Xs,N1), count(X,Ys,N2) => N1=N2.}

\vspace{-.5mm}
\noindent
stating that the input and output of the \tts{sort} program 
have the same multiset of integers.

%



{As an example of a verification problem for a tree manipulating 
program, in Figure~\ref{fig:bstdel} 
we present: (i)~the clauses for \tts{bstdel(X,T1,T2)}, which deletes 
the element \tts{X} from the
binary search tree \tts{T1}, thereby deriving the tree \tts{T2}, and (ii)~a contract 
for \tts{bstdel} (catamorphism \tts{bstree} is shown in Figure~\ref{fig:cataEx3})}.
\vericat proved 
that contract 
by first transforming 
the clauses of Figure~\ref{fig:bstdel} and the following goal that represents the contract for \tts{bstdel}:

\vspace*{-1mm}
\begin{sizepar}{9}{10} 
\begin{verbatim}
    false :- ~(B1=B2), bstree(T1,B1), bstdel(X,T1,T2), bstree(T2,B2).
\end{verbatim} 
\end{sizepar}
\vspace{-2mm}

\vspace{-1mm}

\begin{figure}[ht!]
\begin{sizepar}{8.5}{11}
\begin{verbatim}
bstdel(X,leaf,leaf).
bstdel(X,node(A,leaf,R),R) :- X=A.       
bstdel(X,node(A,L,leaf),L) :- X=A.
bstdel(X,node(A,L,R),node(A1,L,T)) :- X=A & D, delmin(R,T,A1,D).
bstdel(X,node(A,L,R),node(A,L1,R)) :- X<A, bstdel(X,L,L1).
bstdel(X,node(A,L,R),node(A,L,R1)) :- A<X, bstdel(X,R,R1).
delmin(leaf,leaf,M,D) :- M=0 & ~D.       
delmin(node(A,leaf,R),R,M,D) :- M=A & D.
delmin(node(A,node(B,U,V),R),node(A,T,R),M,D) :- D, delmin(node(B,U,V),T,M,D).
:- spec bstdel(X,T1,T2) ==> bstree(T1,B1), bstree(T2,B2) => (B1=B2).
\end{verbatim}
\end{sizepar}
\vspace*{-4mm}
\caption{Deleting an element from a binary search tree: \tts{bstdel} 
and its contract. 
\label{fig:bstdel}}
\vspace{-2mm}
\end{figure}

\begin{table}[!ht]
\centering
\begin{tabular}{|@{\hspace{1mm}} l ||r@{\hspace{2mm}}|r@{\hspace{2mm}}|r@{\hspace{2mm}}| r @{\hspace{2mm}}|}
\hline
{\parbox[center]{34mm}{\hspace{10mm}Program}}       & 
{\parbox[top]{20mm}{\vspace*{-1mm}{\begin{center}~\vericat: \\[-.8mm] \# of proved \\[-.8mm] contracts  \end{center}\vspace*{-1mm}}}} & 
{\parbox[top]{20mm}{\vspace*{-1mm}{\begin{center}~\vericat: \\[-.8mm] ~\textrm{time needed} \\[-.8mm] \textrm{for} \textit{Transf}  \end{center}\vspace*{-1mm}}}} & 
{\parbox[top]{20mm}{\vspace*{-1mm}{\begin{center}~\vericat: \\[-.8mm] \textrm{time needed} \\[-.8mm] \textrm{for} \textit{CheckSat}  \end{center}\vspace*{-1mm}}}} & 
{\parbox[top]{20mm}{\vspace*{-1mm}{\begin{center}~AdtRem: \\[-.8mm] \# of proved \\[-.8mm] contracts  \end{center}\vspace*{-1mm}}}} \\ 
\hline\hline
List Membership \rule{0mm}{3.5mm} & 3              & 4\,410                   & 160 & 3     \\[-.5mm]
List Permutation      & 12                         & 17\,840                  & 1\,150   & 12 \\[-.5mm]
List Concatenation    & 9                          & 14\,430                  & 2\,280   & 4 \\[-.5mm]
Reverse               & 20                         & 27\,790                  & 2\,350  & 12  \\[-.5mm]
Double Reverse        & 6                          & 9\,810                   & 1\,930  & 0 \\[-.5mm]
Reverse w/Accumulator & 6                          & 9\,780                   & 380   & 3   \\[-.5mm]
Bubblesort           & 18                         & 27\,300                  & 1\,270  & 18  \\[-.5mm]
Insertionsort        & 12                         & 17\,670                  & 1\,300   & 11 \\[-.5mm]
Mergesort            & 25                         & 35\,810                  & 2\,060   & 17  \\[-.5mm]
Quicksort (version 1)   & 19                      & 27\,200                  & 3\,770  & 12  \\[-.5mm]  
Quicksort (version 2)   & 18                      & 25\,930                  & 4\,700  & 11  \\[-.5mm]  
Selectionsort        & 20                         & 31\,160                  & 2\,680 & 18   \\[-.5mm]
Treesort             & 10                         & 15\,250                  & 3\,530   & 1  \\[-.5mm]
Binary Search Tree    & 13                         & 21\,330                  & 4\,820   & 3 \\[-.2mm]
\hline\hline
\hspace{10mm}\textrm{Total} \rule{0mm}{3.mm}&	\textrm{191} &	\textrm{285\,710}	& \textrm{32\,380}  &  125 \\
\hline
\end{tabular}
\vspace{-1mm}
\caption{
Contracts proved by the \vericat and  AdtRem tools. Times are in ms.}
\label{tab:exp}
\vspace*{-3mm}
\end{table}

\smallskip
In Table~\ref{tab:exp} we summarize the results of our 
experiments {performed on} an Intel Xeon CPU E5-2640 2.00GHz with 64GB RAM under CentOS.
The columns report the name of the program, the 
number of contracts 
proved by \vericat for each program, and
the total time, in milliseconds, needed for  the\,{\em Transf} and\,{\em CheckSat} steps.
Finally, as a baseline, we report
the number of contracts 
proved by AdtRem~\cite{De&22a}, 
which does not take advantage of the 
user-provided 
contract specifications 
and, instead, 
tries to discover lemmas by using the differential replacement transformation rule.
Our results show that VeriCaT is indeed able to 
exploit the extra information provided by the contracts, and
performs better than previous transformational approaches.

In order to compare the effectiveness of our method with that of other tools,
we have also run solvers such as
AdtInd~\cite{Ya&19}, CVC4 extended with induction~\cite{ReK15}, Eldarica (2.0.6), and SPACER (with Z3 4.8.12)
on the CHC specifications (translated to SMT-LIB format) 
\textit{before} the
application of the\,\textit{Transf} {step}.
Eldarica and SPACER proved the satisfiability of the CHCs for 12 and 1 contracts, respectively,
while the AdtInd and CVC4 did not solve any problem within the time limit of 300~s.
However, it might be the case that better results can be
achieved by those tools by using some 
different encodings of the verification problems.
Finally, we ran the {Stainless} verifier (0.9.1)~\cite{HamzaVK19}
on a few manually encoded specifications of programs
for reversing a list, 
deleting an element from a binary search tree, 
and sorting a list using the Quicksort algorithm.
{Stainless} can prove some, but not all contracts of each of these specifications.
For a more exhaustive comparison we would need an automated translator, 
which is not available yet, between
CHCs and {Stainless} specifications. 
\vspace{-3mm}

	\section{Related Work and Conclusions} 
	\label{sec:RelConcl}

%
%
\vspace*{-2mm}

Many program verifiers are based on Hoare's axiomatic notion of correctness~\cite{Hoa69} 
and have the objective of proving the validity of pre/postconditions. 
Some of those verifiers use SMT solvers as a back-end 
to check verification conditions in various logical theories~\cite{Boogie,Su&11,Lei13,Why3,HamzaVK19}.
In order to deal with properties of programs that manipulate 
ADTs, program verifiers may also be 
enhanced with 
some form of induction (e.g., {Dafny}~\cite{Lei13}), or be based on the unfolding 
of recursive functions (e.g., {Leon}~\cite{Su&11} and {Stainless}~\cite{HamzaVK19}),
or rely on an SMT solver that uses induction, such as the extension of CVC4 
proposed by Reynolds and Kuncak~\citeyearpar{ReK15}. 

Catamorphisms were used 
to define decision procedures for suitable classes of SMT 
formulas~\cite{SuterDK10,PhamGW16}, and
a special form of integer-valued catamorphisms, called {\em type-based norms}, 
were used 
for proving termination of logic programs~\cite{BruynoogheCGGV07} and for
{\em resource analysis}~\cite{AlbertGGM20}.
The main difference of  our approach with respect to these works
is that we transform a set of clauses with catamorphisms 
into a new set of CHCs that act  
on the codomains of the catamorphisms, and in those clauses neither ADTs nor 
catamorphisms are present.

The contracts considered in this paper are similar to 
\texttt{calls} and \texttt{success} user-defined assertions
supported by the Ciao logic programming system~\cite{Hermenegildo&12}.
Those assertions may refer to operational properties of logic 
programs, taking into account the order of execution and the 
extra-logical features of the language, while here we consider only the logical meaning of CHCs
and contracts.
By using abstract interpretation techniques,
the CiaoPP preprocessor~\cite{HermenegildoPBL05} can statically verify 
a wide range of \texttt{calls}/\texttt{success} assertions related to
types, modes, non-failure, determinism, and typed-norms. 
However, CiaoPP suffers from some limitations in checking
the validity of properties expressed by constrained types,
such as the catamorphisms considered {in this paper,
e.g., the properties} of being a sorted list or a binary search tree.

The use of CHCs for program verification has become very popular and many
techniques and tools for translating program verification problems into
satisfiability problems for CHCs have been proposed (see, for instance, the
surveys by Bj{\o}rner et~al.~\citeyearpar{Bj&15}   
and by De Angelis et al.~\citeyearpar{DeAngelisFGHPP21}).
However, as also shown in this paper, in the case of clauses 
with ADT terms, state-of-the-art CHC solvers have some severe limitations
due to the fact that they do not include any proof technique for inductive reasoning on the
ADT structures.
Some approaches to mitigate these limitations include:
(i)~a proof system that combines inductive theorem proving with
CHC solving~\cite{Un&17},
(ii)~lemma generation based on syntax-guided synthesis from user-specified templates~\cite{Ya&19},
(iii)~invariant discovery based on finite tree automata~\cite{KostyukovMF21},
and (iv)~use of suitable abstractions~\cite{GovindSG22}.
%

\smallskip
Transformation-based approaches to the verification of
CHC satisfiability on ADTs have been proposed in recent work~\cite{MoF17,De&18a,KobayashiFG20},
with the aim to avoid the complexity of integrating CHC solving with induction.
The transformational approach 
compares well with induction-based solvers, but
it also shares similar issues for full mechanization, such as the need for 
lemma discovery~\cite{Ya&19,De&22a}.
The transformation technique we have proposed in this paper avoids the
problem of lemma discovery by relying on user-specified contracts and, unlike previous work~\cite{De&18a,De&22a}, it guarantees the termination of the transformation for a large 
class of CHCs.

Our experiments show that the 
novel transformation technique we propose can 
successfully exploit the 
information supplied by the user-provided 
contracts, and
indeed, (i)~it can increase the effectiveness of state-of-the-art CHC solvers
in verifying contracts encoded as CHCs,
and (ii)~performs better than previous transformational approaches based on lemma discovery.

%
%

For future work, we plan to extend the practical applicability of 
our verification method by developing automatic translators to CHCs of
programs and contracts written in the
languages used by verifiers such as Dafny, Stainless, and Why3. 

\vspace*{-3mm}

\section*{Acknowledgments}
\vspace*{-2mm}

The authors warmly thank the anonymous reviewers for their helpful comments and suggestions.
The authors  are members of the INdAM Research Group GNCS.

\vspace*{-5.5mm}




	\renewcommand\thesection{\Alph{section}}
	\setcounter{section}{0}


\titleformat{\section}[hang]{\bfseries}{}{0pt}{\center{#1\quad\thesection}}


	\vspace{-5mm}
	\section{Appendix\!\!}
	\label{app:Init-clauses}
	\noindent 
The clauses derived from clauses~\tts{1--12} of the set {\ttsi{Reverse}} of clauses (see Figure~\ref{fig:RevCHCs}) and 
goal~\tts{14} (that is, the goal representing the contract for \tts{snoc}) are the following ones: 

\vspace{-1mm}
\begin{sizepar}{8}{11}
\begin{verbatim}
T6. new2(A,B,C,D,E,F,G,H,I) :- D=I & I=J & A & B=J & C=(K=>(J>=L & M)) & E & 
         ~F & G=0 & H & M & ~K & L=0.      
T7. new2(A,B,C,D,E,F,G,H,I) :- D=I & D=J & I=K & K=J & L=M & A & B=M & C=(N=>(M>=V & P)) & 
         E=(D=<L & Q) & F & G=L & H=(R=>(L>=S & T)) & (T & Q)=>P, new2(N,V,P,J,Q,R,S,T,K).
T8. false :- A=B & ~((C & D)=>E), new2(F,G,E,B,D,H,I,C,A).
\end{verbatim}
\end{sizepar}
\vspace{-1mm}
\noindent
Clauses \tts{T6--T8} are shown to be satisfiable by the solvers Eldarica and SPACER.

	\vspace{-5mm}
	\section{Appendix\!\!}
	\label{app:Catamorphisms}
	With reference to Section~\ref{sec:Cata}, in this appendix we show that any instance of Schema~(C) 
of Figure~\ref{fig:cataExt}  
can be transformed into an equivalent instance of Schema~(A) of Figure~\ref{fig:cata3},
for any \tts{h}, \tts{f}, and \tts{combine3} functions.
For this transformation, we use the fold/unfold rules~\cite{TaS84,EtG96}.

For the reader's convenience, we recall the definition of \tts{h}, where, for reasons of simplicity, we assume that 
\tts{f} is an instance of Schema~(A) (however, the transformation
can easily be generalized to the case when \tts{f} is an instance of 
the same Schema~(C)):

\vspace{-1mm}
{\small\begin{verbatim}
   h(X,[],Res) :- base3(X,Res).
   h(X,[H|T],Res) :- h(X,T,R), f(X,T,Rf), combine3(X,H,R,Rf,Res).
   f(X,[],Rf) :- base5(X,Rf).
   f(X,[H|T],Rf) :- f(X,T,RT), combine5(X,H,RT,Rf).
\end{verbatim}
}

\vspace{-1mm}
\noindent
We introduce a new predicate \tts{hf} defined as follows:

\vspace{-1mm}
{\small
\begin{verbatim}
   hf(X,L,R1,R2) :- h(X,L,R1), f(X,L,R2).
\end{verbatim}
}

\vspace{-1mm}
\noindent
By unfolding, using the clauses defining \tts{h} and \tts{f}, we get:

{\small\begin{verbatim}
   hf(X,[],R1,R2) :- base3(X,R1), base5(X,R2).
   hf(X,[H|T],R1,R2) :- h(X,T,R), f(X,T,Rf), combine3(X,H,R,Rf,R1),
                        f(X,T,RT), combine5(X,H,RT,R2).
\end{verbatim}
}

\vspace{-1mm}
\noindent
Since \texttt{f} is a functional predicate, we have that \tts{Rf} = \tts{RT}, and we get:

\vspace{-1mm}
{\small\begin{verbatim}
   hf(X,[],R1,R2) :- base3(X,R1), base5(X,R2).
   hf(X,[H|T],R1,R2) :- h(X,T,R), f(X,T,Rf), combine3(X,H,R,Rf,R1),
                        combine5(X,H,Rf,R2). 
\end{verbatim}
}

\vspace{-1mm}
\noindent
By folding, we replace the conjunction `\tts{h(X,T,R), f(X,T,Rf)}' by (a variant of) the head of the definition
of \tts{hf}, and we derive:

\vspace{-1mm}
{\small\begin{verbatim}
   hf(X,[],R1,R2) :- base3(X,R1), base5(X,R2).
   hf(X,[H|T],R1,R2) :- hf(X,T,R,Rf), combine3(X,H,R,Rf,R1), combine5(X,H,Rf,R2).
\end{verbatim}
}

\vspace{-1mm}
\noindent
which is an instance of Schema~(A).
Finally,  we can replace the atom \tts{h(X,L,R1)} in the body of any clause by 
the atom \tts{hf(X,L,R1,R2)} (by first renaming \tts{R2}, if it occurs in the clause).

We leave to the reader to show, in a similar way, that any instance of Schema~(D)
of Figure~\ref{fig:cataExt}  
can be transformed into an equivalent instance of Schema~(B) of Figure~\ref{fig:cata3}. \hfill $\Box$

	\vspace{-5mm}
	\section{Appendix\!\!}
	\label{app:Proofs}
\noindent
{\it Theorem~\ref{thm:contract} $($Correctness of CHC Translation$)$} 
\vspace*{-2mm}
\begin{proof}
By Definition~\ref{def:valid}, we have that contract $K$ 
is valid with respect to a set $P$ of CHCs if and only if 
$M(P) \models \forall (\mathit{pred}(Z) \wedge c \wedge \mathit{Catas} 
\rightarrow d)$. Since~$P$ is a set of definite CHCs and $M(P)$ is its unique 
least $\mathbb D$-model,
$M(P) \models \forall (\mathit{pred}(Z) \wedge c \wedge \mathit{Catas}
\rightarrow d)$ if and only if $P\cup \{\mathit{false} \leftarrow \neg\,d, c, 
\mathit{pred}(Z), \mathit{Catas}\}$ is satisfiable.  \hfill 
\end{proof}

\vspace*{2mm}
\noindent
{\it Theorem~\ref{thm:termination} $($Termination of {Algorithm}~\Cata$)$}
\vspace*{-2mm}
\begin{proof}
Each execution of the $\mathit{Define}$, 
$\mathit{Unfold}$, ${\mathit{Apply}}$-${\mathit{Contracts}}$, and  ${\mathit{Fold}}$ procedures terminates.
The {\bf while}-{\bf do} of~\Cata~terminates if and only if the set $\mathit{Defs}$ of new
definitions introduced during the various iterations cannot grow indefinitely.
A bounded growth of $\mathit{Defs}$ is guaranteed by the fact that, by construction, every definition in $\mathit{Defs}$
is of the form $\mathit{newp}(U) \If c, A, \mathit{Catas}\!_A$ and:
(i)~there is exactly one atom $A$ of the form $p(Z)$, where $p$ is a program predicate and $Z$ is a tuple of distinct variables,
(ii)~for each ADT variable~$Z_{i}$ in~$A$, the conjunction 
$\mathit{Catas}\!_A$ contains occurrences of distinct catamorphism atoms with~$Z_{i}$, and
(iii)~the constraint~$c$ is obtained by a sequence of applications of a widening operator,
and hence this sequence cannot be infinite.  \hfill  
\end{proof}

\vspace*{2mm}
\noindent
{\it Theorem~\ref{thm:soundness-AlgorithmR} $($Soundness of {Algorithm}~\Cata$)$}
\vspace*{-2mm}
\begin{proof}
Let $\mathit{Gs}$ be the set $\{\gamma(K) \mid K\!\!\in\!\!\mathit{Cns}\}$
of goals that translate the contracts in $\mathit{Cns}$.
Thus, by Theorem~\ref{thm:contract}, the contracts in $\mathit{Cns}$
are valid with respect to $P$ if and only if \mbox{$P\cup \mathit{Gs}$} is 
satisfiable.
{Algorithm}~\Cata~can be viewed as a transformation 
of $P\cup \mathit{Gs}$ into $\mathit{TransfCls}$ using 
the following transformation rules presented in the literature~\cite{TaS84,EtG96}:
(i)~{\em definition introduction}, (ii)~{\em unfold}, (iii)~{\em fold}, and (iv)~{\em goal replacement}
(based  on the functionality and totality properties for catamorphisms, and 
on the validity of contracts for auxiliary functions).
In particular, the results presented in recent work~\cite{De&22a},
which applies those transformation rules to the 
proof of 
CHC satisfiability,
guarantee that, if $\mathit{TransfCls}$ is satisfiable, 
then also $P\cup \mathit{Gs}$ is satisfiable. \hfill 
\end{proof}



\begin{thebibliography}{}


\vspace*{-.65mm}\bibitem[Albert et~al., 2020]{AlbertGGM20}
{\sc Albert, E.}, {\sc Genaim, S.}, {\sc Guti{\'{e}}rrez, R.}, {\sc and} {\sc
  Martin{-}Martin, E.}, 2020.
\newblock A transformational approach to resource analysis with typed-norms
  inference.
\newblock {\em Theory Pract. Log. Program.}, {\it 20}, 3, 310--357.


\vspace*{-.65mm}\bibitem[Barnett et~al., 2006]{Boogie}
{\sc Barnett, M.}, {\sc Chang, B.-Y.~E.}, {\sc {De Line}, R.}, {\sc Jacobs,
  B.}, {\sc and} {\sc Leino, K. R.~M.}, 2006.
\newblock Boogie: {A} modular reusable verifier for object-oriented programs.
\newblock 
  {\em Formal Methods for Components and
  Objects}, LNCS 4111, pp. 364--387.
  Springer.
  
\vspace*{-.65mm}\bibitem[Barrett et~al., 2011]{CVC4}
{\sc Barrett, C.}, {\sc Conway, C.~L.}, {\sc Deters, M.}, {\sc Hadarean, L.},
  {\sc Jovanovic, D.}, {\sc King, T.}, {\sc Reynolds, A.}, {\sc and} {\sc
  Tinelli, C.}, 2011.
\newblock {CVC4}.
\newblock 
{\em
  23rd {CAV}}, 
  LNCS 6806, pp. 171--177.
  Springer.

\vspace*{-.65mm}\bibitem[Barrett et~al., 2009]{Ba&09}
{\sc Barrett, C.~W.}, {\sc Sebastiani, R.}, {\sc Seshia, S.~A.}, {\sc and} {\sc Tinelli, C.}
\newblock Satisfiability modulo theories.
\newblock {\em Handbook of Satisfiability}, volume 185
of {\em Frontiers in Artificial Intelligence and Applications}, pp. 825--885.
{IOS} Press.

\vspace*{-.65mm}\bibitem[Bj{\o}rner et~al., 2015]{Bj&15}
{\sc Bj{\o}rner, N.}, {\sc Gurfinkel, A.}, {\sc McMillan, K.~L.}, {\sc and}
  {\sc Rybalchenko, A.},  2015.
\newblock Horn clause solvers for program verification.
\newblock 
  {\em Fields of
  Logic and Computation {\rm{(II)}}}, LNCS
  9300, pp. 24--51. Springer.

\vspace*{-.65mm}\bibitem[Booch and Bryan, 1994]{BoochB94}
{\sc Booch, G.} {\sc and} {\sc Bryan, D.}, 1994.
\newblock {\em Software Engineering with Ada {$($3.} ed.$)$}
\newblock Series in Object-Oriented Software Engineering.
  Benjamin/Cummings.

\vspace*{-.65mm}\bibitem[Bruynooghe et~al., 2007]{BruynoogheCGGV07}
{\sc Bruynooghe, M.}, {\sc Codish, M.}, {\sc Gallagher, J.~P.}, {\sc Genaim,
  S.}, {\sc and} {\sc Vanhoof, W.}, 2007.
\newblock Termination analysis of logic programs through combination of
  type-based norms.
\newblock {\em {ACM} Trans. Program. Lang. Syst.}, {\it 29}, 2, 10--es.


\vspace*{-.65mm}\bibitem[Cousot and Halbwachs, 1978]{CoH78}
{\sc Cousot, P.} {\sc and} {\sc Halbwachs, N.},  1978.
\newblock Automatic discovery of linear restraints among variables of a
  program.
\newblock 
{\em 5th POPL}, pp. 84--96. {ACM}.

\vspace*{-.65mm}\bibitem[{De Angelis} et~al., 2021]{DeAngelisFGHPP21}
{\sc {De Angelis}, E.}, {\sc Fioravanti, F.}, {\sc Gallagher, J.~P.}, {\sc
  Hermenegildo, M.~V.}, {\sc Pettorossi, A.}, {\sc and} {\sc Proietti, M.},
  2021.
\newblock Analysis and transformation of constrained {H}orn clauses for program
  verification.
\newblock {\em Theory Pract. Log. Program.},  pp. 1--69,
\newblock {doi: 10.1017/S1471068421000211}.


\vspace*{-.65mm}\bibitem[{De~Angelis} et~al., 2014]{De&14b}
{\sc {De~Angelis}, E.}, {\sc Fioravanti, F.}, {\sc Pettorossi, A.}, {\sc and}
  {\sc Proietti, M.},  2014.
\newblock {VeriMAP}: {A} tool for verifying programs through transformations.
\newblock {\em 20th TACAS}, LNCS 8413,
  pp. 568--574. Springer.

%
\vspace*{-.65mm}\bibitem[{De Angelis} et~al., 2018]{De&18a}
{\sc {De Angelis}, E.}, {\sc Fioravanti, F.}, {\sc Pettorossi, A.}, {\sc and}
  {\sc Proietti, M.}, 2018.
\newblock Solving {H}orn clauses on inductive data types without induction.
\newblock {\em Theory Pract. Log. Program.}, {\it 18}, 3-4,
  452--469.

\vspace*{-.65mm}\bibitem[{De Angelis} et~al., 2022]{De&22a}
{\sc {De Angelis}, E.}, {\sc Fioravanti, F.}, {\sc Pettorossi, A.}, {\sc and}
{\sc Proietti, M.} 2022.
\newblock Satisfiability of constrained {H}orn clauses on algebraic data types:
{A} transformation-based approach.
\newblock {\em J. Log. Comput.}, {\it 32}, 2, 402--442.


\vspace*{-.65mm}\bibitem[de~Moura and Bj{\o}rner, 2008]{DeB08}
{\sc de~Moura, L.~M.} {\sc and} {\sc Bj{\o}rner, N.}, 2008.
\newblock Z3: {A}n efficient {SMT} solver.
\newblock 
{\em 14th {TACAS}}, LNCS
  4963, pp. 337--340. Springer.


\vspace*{-.65mm}\bibitem[Etalle and Gabbrielli, 1996]{EtG96}
{\sc Etalle, S.} {\sc and} {\sc Gabbrielli, M.}, 1996.
\newblock Trans\-form\-ations of {CLP} modules.
\newblock {\em Theor. Comput. Sci.}, {\it 166}, 101--146.

\vspace*{-.65mm}\bibitem[Filli{\^{a}}tre and Paskevich, 2013]{Why3}
{\sc Filli{\^{a}}tre, J.-C.} {\sc and} {\sc Paskevich, A.}, 2013.
\newblock Why3 - {W}here programs meet provers.
\newblock 
{\em
  22nd {ESOP}},  
  LNCS~7792, pp. 125--128. Springer.


\vspace*{-.65mm}\bibitem[{Govind V. K.} et~al., 2022]{GovindSG22}
{\sc {Govind V. K.}, H.}, {\sc Shoham, S.}, {\sc and} {\sc Gurfinkel, A.} 2022.
\newblock Solving constrained {H}orn clauses modulo algebraic data types and
recursive functions.
\newblock {\em Proc. {ACM} Program. Lang.}, {\it 6}, {POPL}, 1--29.

\vspace*{-.65mm}\bibitem[Grebenshchikov et~al., 2012]{Gr&12}
{\sc Grebenshchikov, S.}, {\sc Lopes, N.~P.}, {\sc Popeea, C.}, {\sc and} {\sc
  Rybalchenko, A.}, 2012.
\newblock Synthesizing software verifiers from proof rules.
\newblock 
{\em 33rd
PLDI}, pp. 405--416.


\vspace*{-.65mm}\bibitem[Hamza et~al., 2019]{HamzaVK19}
{\sc Hamza, J.}, {\sc Voirol, N.}, {\sc and} {\sc Kuncak, V.}, 2019.
\newblock System {FR:} Formalized foundations for the Stainless verifier.
\newblock {\em {ACM} Program. Lang.}, {\it 3}, {OOPSLA}, 166:1--166:30.

\vspace*{-.65mm}\bibitem[Hermenegildo et~al., 2012]{Hermenegildo&12}
{\sc Hermenegildo, M.}, {\sc Bueno, F.}, {\sc Carro, M.}, {\sc
  L{\'o}pez-Garc{\'i}a, P.}, {\sc Mera, E.}, {\sc Morales, J.~F.}, {\sc and}
  {\sc Puebla, G.}, 2012.
\newblock {A}n overview of {C}iao and its design philosophy.
\newblock {\em Theor. Pract. Logic Program.}, {\it 12}, 1--2, 219--252.

\vspace*{-.65mm}\bibitem[Hermenegildo et~al., 2005]{HermenegildoPBL05}
{\sc Hermenegildo, M.~V.}, {\sc Puebla, G.}, {\sc Bueno, F.}, {\sc and} {\sc
  L{\'{o}}pez{-}Garc{\'{\i}}a, P.}, 2005.
\newblock Integrated program debugging, verification, and optimization using
  abstract interpretation (and the {C}iao system preprocessor).
\newblock {\em Sci. Comput. Program.}, {\it 58}, 1-2, 115--140.

\vspace*{-.65mm}\bibitem[Hinze et~al., 2013]{HinzeWG13}
{\sc Hinze, R.}, {\sc Wu, N.}, {\sc and} {\sc Gibbons, J.}
\newblock Unifying structured recursion schemes.
\newblock ICFP 2013, pp. 209--220. {ACM}.

\newcommand{\bsp}{\hspace*{-.5pt}}
\newcommand{\wsp}{\hspace*{1pt}}
\vspace*{-.65mm}\bibitem[Hoare, 1969]{Hoa69}
{\sc Hoare,\,\bsp C\bsp.\bsp A\bsp.\bsp R\bsp.\bsp},\bsp 1969.
\newblock \!An\,{A}xiomatic\,{B}asis\,for\,{C}omputer\,{P}rogramming.\,
\newblock {\em \!\!\! CACM},\,\bsp{\it\!12},10,\wsp576--580,\wsp583.


\vspace*{-.65mm}\bibitem[Hojjat and R{\"{u}}mmer, 2018]{HoR18}
{\sc Hojjat, H.} {\sc and} {\sc R{\"{u}}mmer, P.}, 2018.
\newblock The {ELDARICA} {H}orn solver.
\newblock 
{\em
  Formal Methods in Computer Aided Design, \mbox{FMCAD}}, pp. 1--7.
  {IEEE}.

\vspace*{-.65mm}\bibitem[Jaffar and Maher, 1994]{JaM94}
{\sc Jaffar, J.} {\sc and} {\sc Maher, M.}, 1994.
\newblock Constraint logic programming: {A} survey.
\newblock {\em Journal of Logic Programming}, {\it 19/20}, 503--581.


\vspace*{-.65mm}\bibitem[Kobayashi et~al., 2020]{KobayashiFG20}
{\sc Kobayashi, N.}, {\sc Fedyukovich, G.}, {\sc and} {\sc Gupta, A.}, 2020.
\newblock Fold/unfold transformations for fixpoint logic.
\newblock 
{\em 
  26th {TACAS}}, 
  LNCS 12079, pp. 195--214. Springer.

\vspace*{-.65mm}\bibitem[Komuravelli et~al., 2014]{Ko&14}
{\sc Komuravelli, A.}, {\sc Gurfinkel, A.}, {\sc and} {\sc Chaki, S.}, 2014.
\newblock {SMT}-based model checking for recursive programs.
\newblock 
{\em 26th {CAV}}, LNCS 8559,
  pp. 17--34. Springer.

\vspace*{-.65mm}\bibitem[Kostyukov et~al., 2021]{KostyukovMF21}
{\sc Kostyukov, Y.}, {\sc Mordvinov, D.}, {\sc and} {\sc Fedyukovich, G.}, 2021.
\newblock Beyond the elementary representations of program invariants over
  algebraic data types.
\newblock 
{\em
  42nd {PLDI}},
pp. 451--465. {ACM}.

\vspace*{-.65mm}\bibitem[Leino, 2013]{Lei13}
{\sc Leino, K.~R.~M.},  2013.
\newblock Developing verified programs with {D}afny. 
\newblock 
{\em Intl. Conf. on Software Engineering}, pp.
  1488--1490. IEEE Press.

\vspace*{-.65mm}\bibitem[Meijer et~al., 1991]{MeijerFP91}
{\sc Meijer, E.}, {\sc Fokkinga, M.~M.}, {\sc and} {\sc Paterson, R.}, 1991.
\newblock Functional programming with bananas, lenses, envelopes and barbed
  wire.
\newblock 
{\em 5th {ACM} Conf. on Functional Programming Languages
  and Computer Architecture}, 
  LNCS 523, pp. 124--144.
  Springer.

%

\vspace*{-.65mm}\bibitem[Meyer, 1992]{Meyer92}
{\sc Meyer, B.}, 1992.
\newblock Applying ``Design by Contract''.
\newblock {\em Computer}, {\it 25}, 10, 40--51.


\vspace*{-.65mm}\bibitem[Mordvinov and Fedyukovich, 2017]{MoF17}
{\sc Mordvinov, D.} {\sc and} {\sc Fedyukovich, G.}, 2017.
\newblock Synchronizing constrained {H}orn clauses.
\newblock 
\mbox{\em LPAR-21},
  {\em EPiC Series in Computing} vol.\,46, pp. 338--355. EasyChair.

\vspace*{-.65mm}\bibitem[Odersky et~al., 2011]{OderskySV11}
{\sc Odersky, M.}, {\sc Spoon, L.}, {\sc and} {\sc Venners, B.}, 2011.
\newblock {\em Programming in {S}cala: {A} Comprehensive Step-by-Step Guide}.
\newblock Artima, Sunnyvale, CA, USA, 2nd Edition. 


\vspace*{-.65mm}\bibitem[Pham et~al., 2016]{PhamGW16}
{\sc Pham, T.}, {\sc Gacek, A.}, {\sc and} {\sc Whalen, M.~W.}, 2016.
\newblock Reasoning about algebraic data types with abstractions.
\newblock {\em J. Autom. Reason.}, {\it 57}, 4, 281--318.

\vspace*{-.65mm}\bibitem[Reynolds and Kuncak, 2015]{ReK15}
{\sc Reynolds, A.} {\sc and} {\sc Kuncak, V.}, 2015.
\newblock Induction for {SMT} solvers.
\newblock 
  {\em 16th {VMCAI}}, LNCS 8931, pp. 80--98. Springer.

\vspace*{-.65mm}\bibitem[Suter et~al., 2010]{SuterDK10}
{\sc Suter, P.}, {\sc Dotta, M.}, {\sc and} {\sc Kuncak, V.}, 2010
\newblock Decision procedures for algebraic data types with abstractions.
\newblock 
  {\em 37th 
  {POPL}},  
  pp. 199--210. {ACM}.

\vspace*{-.65mm}\bibitem[Suter et~al., 2011]{Su&11}
{\sc Suter, P.}, {\sc K\"{o}ksal, A.~S.}, {\sc and} {\sc Kuncak, V.}, 2011.
\newblock Satisfiability modulo recursive programs.
\newblock 
{\em 18th {SAS}}, LNCS~6887, 
pp. 298--315. Springer.

\vspace*{-.65mm}\bibitem[Tamaki and Sato, 1984]{TaS84}
{\sc Tamaki, H.} {\sc and} {\sc Sato, T.}, 1984.
\newblock Unfold/fold trans\-form\-ation of logic pro\-grams.
\newblock 
{\em
  2nd~ICLP}, pp. 127--138,  Uppsala University, 
  Sweden.
  

\vspace*{-.65mm}\bibitem[Unno et~al., 2017]{Un&17}
{\sc Unno, H.}, {\sc Torii, S.}, {\sc and} {\sc Sakamoto, H.}, 2017.
\newblock Automating induction for solving {H}orn clauses.
\newblock 
{\em 29th
  {CAV}}, 
  LNCS 10427, pp. 571--591. Springer.


\vspace*{-.65mm}\bibitem[Yang et~al., 2019]{Ya&19}
{\sc Yang, W.}, {\sc Fedyukovich, G.}, {\sc and} {\sc Gupta, A.}, 2019.
\newblock Lemma synthesis for automating induction over algebraic data types.
\newblock 
{\em 25th
  {CP}},
  LNCS 11802, pp. 600--617. Springer.
\end{thebibliography}
\end{document}